\documentclass[11pt,a4paper]{article}
\usepackage{amssymb,amsmath,graphicx,subcaption,hyperref,cite}

\usepackage[utf8]{inputenc}
\usepackage[english]{babel}
\begin{document}
\begin{center}
{\Large\bf Effects of random fields on the reversal of magnetisation of Ising ferromagnet}
\end{center}
\vskip 1 cm
\begin{center} 
Moumita Naskar$^\ast$ and Muktish Acharyya$^\#$

\textit{Department of Physics, Presidency University, 86/1 College Street, 
Kolkata-700073, India} 

\vskip 0.2 cm
\textit{Email$^\ast$: moumita1.rs@presiuniv.ac.in}
ordered 
\textit{Email$^\#$: muktish.physics@presiuniv.ac.in}
\end{center}
\vspace {1.0 cm}

\noindent {\bf Abstract:} 
We have studied the reversal time of the magnetisation in two dimensional Ising ferromagnet in the presence of 
externally applied uniform magnetic field using Monte Carlo simulation based on Metropolis single spin 
flip algorithm. Then we have investigated the change in reversal time due to the presence of quenched random field in 
addition to the uniform magnetic field. We report the results of statistical distribution of reversal 
times in the presence of three different types of the distributions (namely, uniform, 
bimodal and normal) of random fields and compared the results with those obtained for uniform field only. 
We have observed that the reversal time decreases due to the presence of any kind (of distribution) of the random 
fields. The metastable volume fraction is observed to follow the Avrami's law. Dependence of reversal times on temperature and different widths of the distributions of random fields are also reported. 
We have also checked whether the system obeyed Becker-D\"{o}ring theory of classical nucleation in presence of additional random 
field and tried to investigate the range of the width of the distribution of random field. For larger width of the distribution of random field, the system fails to show
the reversal via the nucleation of a single droplet (for small values of uniform
field only). The possible reason is analysed.
\vspace {2.0 cm}


\newpage
\noindent{\Large\bf I. Introduction}
\vspace {0.5 cm}

The field-induced reversal mechanism of the magnetisation, in materials has been the subject of considerable 
interest since many decades. A ferromagnetic sample at temperature $T < T_c$ (Curie temperature) has a 
magnetic moment along a particular direction in absence of any applied external field. Now if a weak magnetic 
field is applied in the opposite direction of that of the net magnetic moment, then the sample would try to align along the 
direction of field. The change in direction of this net magnetic moment, due to applied 
field in opposite direction, is known as \textit{reversal of magnetisation}. The time required for this reversal of magnetisation is 
called \textit{reversal time}.

Magnetic materials are used in wide variety of devices. In some cases, magnetic materials need to change the 
direction of magnetisation, in the presence of very small magnetic field.  Whereas in some other cases, direction of magnetisation is required 
to keep as stable as possible, against both of the magnetic and thermal noises \cite{vogel}. Let’s take an example of a particular 
device, namely, the magnetic storage media which are widely used in our modern life to store the information. The 
informations have to be written on magnetic storage media in the form of small magnetic grains. Now for faster recording
of the data, the tiny magnetic grains should respond quickly to the external field. On the other hand, for the storage of higher 
longevity of the device, the tiny magnetic grains should be as stable as possible against any kind of effective noise. So, in the practical situation, a compromise   between these two cases \cite{vogel} \cite{memory}, is extremely important. For
 theoretical research, it is quite important 
 to investigate the changes in reversal times of magnetisation in various environment,
 namely thermal noise, quenched disorder etc .

Extensive computer simulation and experimental work, on the decay of metastable state and nucleation are done in last 
few decades. Kinetics of nucleation, particularly rate of nucleation of crystalline droplets in the solid-melt 
systems was studied \cite{mgrant}. Lifetimes of metastable states in kinetic Ising ferromagnets are studied 
by droplet theory to investigate their dependence on applied field and system size \cite{tomita}. Extensive simulational research on 
nucleation in different dimensions $d$ ($d=2,3,4$) has been done using heat-bath dynamics, to verify the results as predicted by
the classical nucleation theory of Becker-D\"{o}ring   
\cite{stauffer}. The  dynamic magnetization-reversal transition in pure Ising systems 
under a pulsed field was also studied 
 \cite{misra}. An investigation about the thermally activated magnetisation switching of small ferromagnetic particles driven by an external magnetic field has been carried out and a crossover from 
{\it coherent rotation} to {\it nucleation} for a classical anisotropic Heisenberg model, has been reported
\cite{hinzke}. The rates of growth and decay of the clusters of different sizes, have been studied \cite{vehkamaki} as functions of external 
field and temperature. The rate of nucleation of critical nuclei, its expansion speed
and the corresponding changes of free energy is well connected and is described by
famous Kolmogorov-Johnson-Mehl-Avrami law\cite{kolmogorov,johnson,avrami}.

Dynamics of magnetization reversal in models of magnetic nanoparticles and ultra-thin films have also been reported \cite{novotny}. Simulational result of magnetisation switching 
in the system of nanoparticles was reported \cite{hinzke2001}. Asymmetric reversal modes in ferromagnetic/ antiferromagnetic multilayers was also studied \cite{usadel}.  Distribution of nucleation times, in the system showing Brownian type dynamics,
has been described by classical nucleation theory \cite{barkema}. Decay of metastable state in a model for the 
catalytic oxidation of CO was studied \cite{machado}. Non-equilibrium magnetisation reversal in kinetic Ising 
ferromagnet driven by a periodic impulsive magnetic field in the meanfield approximation was investigated 
\cite{ma2010}. The heat assisted magnetisation reversal in ultra thin films for ultra-high-density 
information recording has been investigated \cite{deskins}. In a recent paper\cite{ma2014}, nucleation time was observed to 
increase in the presence of magnetic field spreading over the space in time as compared to that in static field. Linear reversal mechanism in FePt grains has been simulated using atomistic spin dynamics, 
parameterized from ab-initio  calculation \cite{ellis}. Very recently, the magnetisation reversal in Ising 
ferromagnet by a field having spatial gradient was studied \cite{adhar}. In continuation, investigation has been 
done with simultaneous presence of field gradient and thermal gradient and a condition of the marginal competition (between field gradient and thermal gradient) 
has been reported \cite{rdutta}. 

Since the thermal fluctuation plays an important role to transit the system
from an ordered state to a disordered state via the spin flip mechanism (in the
case of normal ferro-para phase transition in Ising model), it would be quite
inquisitive effort to know whether any other kind of quenched disorder 
would make the transition. The random field Ising model (RFIM) is a prototype
of such studies.
The ferro-para phase transition
by randomly quenched magnetic field is a remarkable discovery in this field
\cite{skma}. The main result is that when the order parameter has a continuous 
symmetry, the ordered phase is unstable against an arbitraryly weak random field
below four dimensions ($d < 4$). Later, it was shown\cite{fisher} that three dimensional random field Ising model could have spontaneous magnetisation at low
temperatures. The studies in estimating the critical exponent of the phase transition in RFIM have been done\cite{aharony}. The critical exponents of a phase transition
in a $d$ dimensional system with short range interaction and a random quenched 
field are equivalent as those of a $(d-2)$ dimensional pure system. However, in a recent study \cite{fytasprl16,fytasrev} by high statistics simulation in four dimensional
RFIM, indicates that in four dimensions the dimensional reduction as predicted
by the perturbative renormalization group study \cite{skma} does not hold good and three independent critical exponents are needed to describe the transition. Recently,
it is reported \cite{fytasprl19} that the  correlation function in a $d$ dimensional disordered system
has an equivalent measure of that in a $(d-2)$ dimensional clean system. But the
supersymmetric prediction are applicable upto $d=5$ (as checked through high statistics
simulation of zero temperature RFIM) and fail to describe the results in $d=4$.

What will happen to the reversal time in the ferromagnetic system in the presence of
disorder ? This is a pertinent question and should be addressed to the research of
the magnetisation switching phenomena. In this context, we found that the reversal
has been studied\cite{kolesik} in a disordered Ising ferromagnet where the boundary sites experiences
the positive and negative fields. Heterogeneous nucleation has been studied
\cite{scheifele} in two
dimensional Ising model where the impurities were placed on a line of fixed points.

As far as the knowledge of the authors is concerned, the study of magnetisation 
reversal has not yet been carried out for randomly quenched (site dependent but time independent) magnetic field. The reversal in the  random field Ising model, has not
been studied so far. \textit{How will the reversal time be affected if the system experiences a quenched random magnetic 
field? Would it help to reduce the reversal time?} Our main goal of the following work is to address this question keeping in mind that
the magnetic reversal time of a ferromagnetic film plays the key role in the speed of recording. The high speed 
magnetic recording device is extremely important in modern technology (e.g., data storage). \textit{The main 
objective is to reduce the reversal time which produces faster recording.}

Below the Curie temperature $T_C$ the ferromagnetic system is in either of two stable states: an ordered where all the spins are 
either up (acquire positive magnetisation) or all the spins are down (acquire negative magnetisation). Now what happens if 
a magnetic field is applied to the system at $T < T_C$ , in the opposite direction of that of the net initial magnetisation? It is observed that the system remains in a metastable state for certain period of time and then gradually reaches stable equilibrium state (magnetisation directed towards the direction of the applied magnetic field) through 
the decay of this metastable state \cite{gunton} \cite{vehkamaki_book}. It is also observed that the system escapes 
from that metastable state through homogeneous nucleation. The dynamical and statistical characteristics of this 
nucleation process are explained by Classical nucleation theory \cite{becker}. A commonly used simplest system for 
studying the nucleation phenomena, is the well-known Ising model.

We have organised the paper in the following format: in the section II, we have
reviewed briefly the classical nucleation theory. The Ising model, we
considered here, in the presence of magnetic field and the method of Monte
Carlo simulation are described in section III. The simulational results and 
analysis are reported in section IV. The paper ends with a brief summary in section V.

\vskip 0.5 cm


\noindent {\Large {\bf II. Classical Nucleation Theory:}}

\vskip 0.5cm

\begin{figure}[htb]
\begin{center}
\resizebox{10cm}{!}{\includegraphics{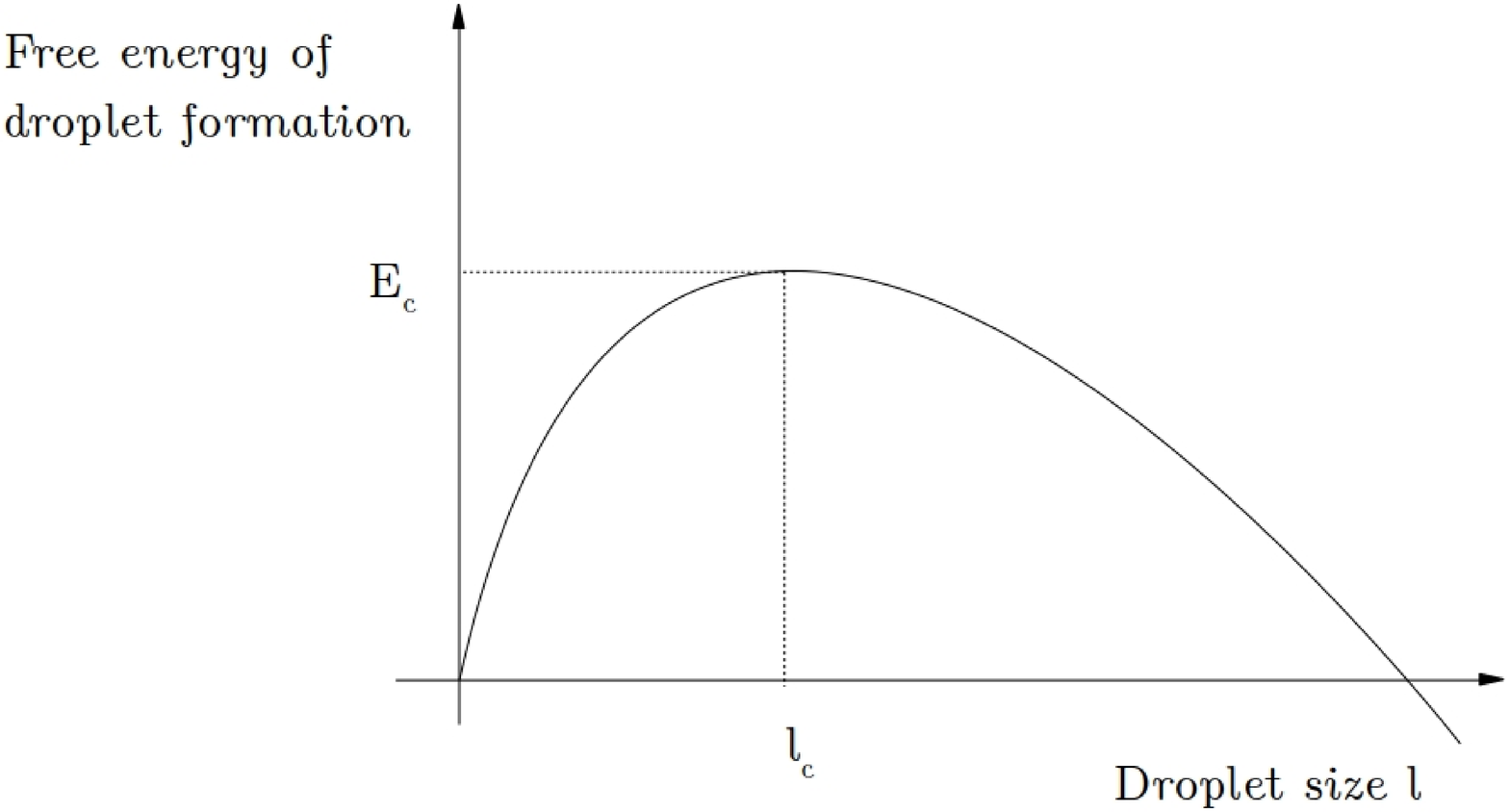}}
\caption{Schematic variation of the free energy (for droplet formation) with size of the droplet}
\label{free_energy}
\end{center}
\end{figure}

If we apply a positive external field, to a ferromagnetic
(spin-1/2 Ising) system at nonzero temperature,
then the small droplets of down spins are dispersed in the background of up spins. But now if a negative external 
field is applied, it is observed that the system reaches the equilibrium state through the decay of a 
initial metastable state. What is happening actually to this metastable state? The number of droplets of down spins of size $l$ 
($l$ number of down spins) is given by Boltzmann factor:
\begin{equation}
n_l = Ne^{-\beta E_l}
\end{equation}
where $\beta=1/k_B T$ and $E_l$ is the free energy of formation of a droplet of size $l$ and N is the normalization 
factor. The classical assumption is that $E_l$ is the sum of the two energies, a bulk energy and a surface energy. The 
bulk energy corresponds to the energy required to flip $l$ number of spins in a field $h$ i.e. equals to $2hl$. The second term i.e. 
the surface energy expresses the energy associated with the surface tension $\sigma$ of the droplet. This is the energy of formation of the boundary of the domain wall. Classical nucleation 
theory\cite{becker,gunton} assumes that the droplets are mostly spherical like in $d$-dimensional space. Thus, 
\begin{subequations}\label{eq:2}
\begin{align}
\label{eq:2a}
E_l=2hl+\sigma l^{(d-1)/d}
\end{align}

\noindent Where $\sigma$ is a prefactor (positive) involving the ferromagnetic interaction 
strength and the constant (combination of Gamma function and some powers of $\pi$ appears in
the calculation of surface area of a $d$ dimensional hypersphere). The radius of
a droplet of size $l$ will be proportional to $l^{1/d}$. The surface area will be
proportional to $l^{(d-1)/d}$.
 Now, if the field is negative then

\begin{align}
\label{eq:2b}
E_l=-2hl+\sigma l^{(d-1)/d}
\end{align}
\end{subequations}
Since the bulk energy is negative and the surface energy is positive,
 there is a competition between the bulk and the 
surface terms (Fig-\ref{free_energy}) as the size of the droplet ($l$) increases. 
The surface term dominates for small $l$ while the bulk term dominates for large $l$. As a consequence there must 
be a critical size of the droplet $l_c$ (with a critical radius $R_c$ of droplet) for which free energy is maximum. In the vicinity of the maximum of the free energy an
interesting dynamics is observed.  If the droplet is supercritical
($l>l_c$), it will grow in size to lower the free energy. In contrast, the
subcritical ($l<l_c$) droplets are found to shrink. As a result, the subcritical
droplets will not take part in the process of reversal.
The supercritical droplets will grow and eventually engulf the whole system producing magnetisation reversal. The transition of the system from metastable phase to a stable phase by uncontrollable growth of spin droplets is known as \textit{nucleation}. The critical size $l_c$ of the droplets can be calculated by maximizing
(differentiating $E_l$ with respect to $l$ in the equation-2b) the free energy and
found as:

\begin{equation}
l_c=\Bigg(\frac{\sigma(d-1)}{2d|h|}\Bigg)^d
\end{equation}
So the maximum free energy for the formation of the droplets is,
\begin{equation}
E_l \Big |_{l=l_C}=E_c=\frac{K_d\sigma^d}{h^{d-1}}
\end{equation}
where $K_d$ is the d dependent constant term. The number of supercritical droplets 
having droplet size $l_c$ is,
\begin{equation}
n_c \sim exp\Bigg(-\frac{E_c}{k_B T}\Bigg) \sim exp\Bigg(-\frac
{K_d \sigma^d}{k_BTh^{d-1}}\Bigg)
\end{equation}

\noindent The \textit{Becker-D\"{o}ring theory}\cite{becker,gunton} explains the behaviour of metastability by the kinetics of cluster 
(droplets of spins) formation. Basic assumption of this theory is that the time evolution of number of droplets is 
only due to an evaporation-condensation mechanism in which a droplet of size $l$ loses or gains a single spin. 
According to this theory, the nucleation rate i.e. the number of droplets formed per unit time and volume, $I$ is proportional to $n_c$
\begin{equation}
I=I_0 e^{-E_c/k_BT}
\end{equation}
$I_0$ is the rate prefactor. The nucleation rate depends very strongly on the exponential term. When 
applied field is very weak then only one supercritical droplet form at a certain time step and it grows with time 
and covers whole system causing reversal of magnetisation. This is called nucleation regime. The time taken by the system to achieve 
magnetisation reversal is called nucleation time. The nucleation time is simply inversely proportional to the 
nucleation rate $I$ derived by Becker-D\"{o}ring theory, 
\begin{subequations}\label{eq:7}
\begin{align}
\label{eq:7a}
\tau_{(nr)} \sim I^{-1} \sim exp\Bigg(\frac{K_d\sigma^d}
{k_BTh^{d-1}}\Bigg)
\end{align}

\noindent So in nucleation regime, for a fixed sample at a fixed temperature,

\begin{align}
\label{eq:7b}
log (\tau_{(nr)}) \sim \frac{1}{h^{d-1}}
\end{align}
\end{subequations}
Now as we increase the value of the field, after certain range we observe a different scenario of magnetisation reversal. 
Here the reversal does not occur through the formation of one single supercritical droplet. Instead, many such 
supercritical droplets grow simultaneously. Outer parts of the droplets touch with each other by forming a bridge 
whose width increases with time. Thus, smaller droplets form bigger droplets through coalescence and span the whole 
system. Radius($\sim l^{\frac{1}{d}}$) of such supercritical droplet grows linearly with time (t) and so the number 
of spins ($l$) will grow as $t^d$. It is very clear that the rate of change in magnetisation is proportional to the 
change in droplet size i.e. number of spins in a droplet 
$$\frac{dm}{dt} \sim \frac{dl}{dt} \sim It^d$$
$$\Longrightarrow \int_{m_1}^{m_2} dm \sim \int_{0}^{\tau_{(cr)}} It^d dt$$
\begin{equation}
\Longrightarrow m_2-m_1 = \Delta m \sim I . \frac{\tau_{(cr)}^{d+1}}{d+1}
\end{equation}
where $\tau_{(cr)}$ is the reversal time at coalescence regime, $m_1$ is the magnetisation at initial stable state 
and $m_2$ is the magnetisation after reversal. So for a fixed change in magnetisation $\Delta m$ reversal time is
\begin{subequations}\label{eq:9}
\begin{align}
\label{eq:9a}
\Longrightarrow \tau_{(cr)} \sim I^{-\frac{1}{d+1}} \sim exp\Bigg(\frac{K_d\sigma^d}
{k_BT(d+1)h^{d-1}}\Bigg)
\end{align}

\noindent So in coalescence regime \cite{stauffer}, for a fixed sample at a fixed temperature,
case for randomly distributed field the
\begin{align}
\label{eq:9b}
log (\tau_{(cr)}) \sim \frac{1}{(d+1) h^{d-1}}
\end{align}
\end{subequations}
So if we plot the logarithm of the reversal time against $1/h$, it would be a fair straight line. The slope of the straight line, in coalescence regime, will be almost  
$\frac{1}{d+1}$ of the slope observed in the nucleation regime. In the last part of our following work we will check whether these 
above relations between reversal time and $1/h$ are also valid in presence of random fields.   
\vskip 2 cm


\noindent {\Large\bf III. Description of the Model and Simulation Scheme}
\vskip 0.5 cm
The Hamiltonian of the pure Ising ferromagnet is represented by,
\begin{subequations}
\label{eq:10}
\begin{align}
\label{eq:10a}
H=-J\sum_{<i,j>}S_iS_j - h_0 \sum_{i}S_i
\end{align}

\noindent where $S_i=\pm1$ are the Ising spins i.e. having two discrete spin states (spin moment axis either pointing up (+1) or 
down (-1)) only. First term in the Hamiltonian represents interaction between nearest neighbour spins with uniform ferromagnetic interaction strength $J(>0)$. Positive $J$ denotes that spins try to align parallel to each other by this cooperative interaction. Second term in the Hamiltonian represents the Zeeman energy, involved in the 
interaction of each spin with the applied uniform external magnetic field
$h_0$. Now the Hamiltonian for the impure Ising ferromagnetic system is represented by,

\begin{align}
\label{eq:10b}
H=-J\sum_{<i,j>}S_i S_j - \sum_{i}h_i S_i
\end{align}
\end{subequations}
where $h_i $ is the quenched (site dependent but time independent) random field which has been used to model the 
effect of impurity in the system. So the form of this field is simply, 
\begin{equation}
h_i=h_0+h_r
\end{equation}
where $h_0$ is the externally applied uniform field and $h_r$ is the site dependent random field whose mean value is 
set to zero ($<h_r>=0$). So that $<h_i>=h_0+<h_r>=h_0$. The mean value of $h_i$ remains same as $h_0$. Thus we can study how does the system's behaviour get affected by a distribution of random field.  We have considered here three different kinds of distributions of random field. The field is measured in the unit of $J$. 

a) \textit{Bimodal distribution:} Having probability distribution, 
\begin{subequations}\label{eq:12}
\begin{align}
\label{eq:12a}
	P_b(r)=0.5\delta(h_r-w)+0.5\delta(h_r+w)
\end{align}

\noindent where $\delta$ represents the Dirac Delta function. This distribution implies that, at 50 $\%$ lattice sites the field is $h_r=+w$ and at another 50$\%$ lattice sites the field is $h_r=-w$ which are randomly oriented into the lattice sites. 

b) \textit{Uniform distribution:} Having probability distribution,
\begin{align}
\label{eq:12b}
	P_u(r)=\frac{1}{n}
\end{align}

\noindent where $n$ is the number of values of field in this range $-w$ to $+w$. So all the values of field between this range are randomly present on the lattice sites with equal proportion.

c) \textit{Gaussian distribution:} Having probability distribution
(obtained by using Box-Muller algorithm\cite{box}) with standard deviation $\sigma(=2w$  considered throughout our work)
\begin{align}
\label{eq:12c}
	P_n(r)=\frac{1}{\sqrt{2\pi\sigma^2}}e^{-\frac{h_r^2}{2\sigma^2}}
\end{align}
\end{subequations}

In above distributions, we have denoted the \textbf{width/ strength of random field}  as '$w$' where values of the random field ($h_r$) are distributed between $h_r=-w$ to $h_r=w$. That means the value of total random field varies from $h_i= h_0-w$ to $h_i= h_0+w$. So the actual width of the distribution of random field ($h_i$) would be $2w$. But we have denoted it simply as $w$.

We have simulated \cite{binder} the two dimensional ferromagnetic Ising square lattice of size L$\times$L with periodic boundary condition applied in 
 both directions. Initially we have considered the system to be in perfectly ordered state where all the spins are pointing up $S_i=1$ $\forall$ i. In our simulation, we have randomly selected a spin out of total $L^2$. That spin has been updated according
 to the Metropolis rate\cite{binder,metro}.
\begin{equation}
 P(S_i\to-S_i) = Min(1,e^{- \frac{\Delta E}{k_B T}})
\end{equation}

where $K_B$ is the Boltzmann constant and $T$ is the temperature of the system
which is measured in the unt of $J/K$. The energies
are measured in the unit of $J$.
The change in energy due to spin flip is $\Delta E=E_f-E_i$.
We have implemented the algorithm as,

i) We have chosen any lattice site randomly and calculated the energy $E_i$ of the system.  Then we have flipped 
the spin at that site and calculated the energy $E_f$. So we got the above exponential 
factor $e^{- \frac{\Delta E}{k_B T}}$.
 
ii) Now we have called any random number uniformly distributed between 0 to 1 and checked that if the above exponential factor is greater than the random number then this flipped state of the system is accepted otherwise the chosen spin is considered to be in its initial unflipped state.

So this was for one single update. $L\times L$ numbers of such random updates are considered as the unit of time in this simulation and called Monte Carlo step per spin
(MCSS). The magnetisation of the system is determined by
\begin{equation}
m(t)=\frac{1}{L^2} \sum_{i}^{L^2} S_i  
\end{equation}

Here, we have chosen $L=300$ everywhere in the simulation process except in the case where we have studied Becker-D\"{o}ring analysis (we have simulated here the lattice of size $L=100$) for affordable computational time.

\vskip 2 cm
\noindent {\large\bf IV. Results and Discussions} 
\vskip 0.5 cm

\begin{figure}[h!]
	\begin{subfigure}[b]{0.5\textwidth}
		\includegraphics[width=0.8\textwidth,angle=-90]{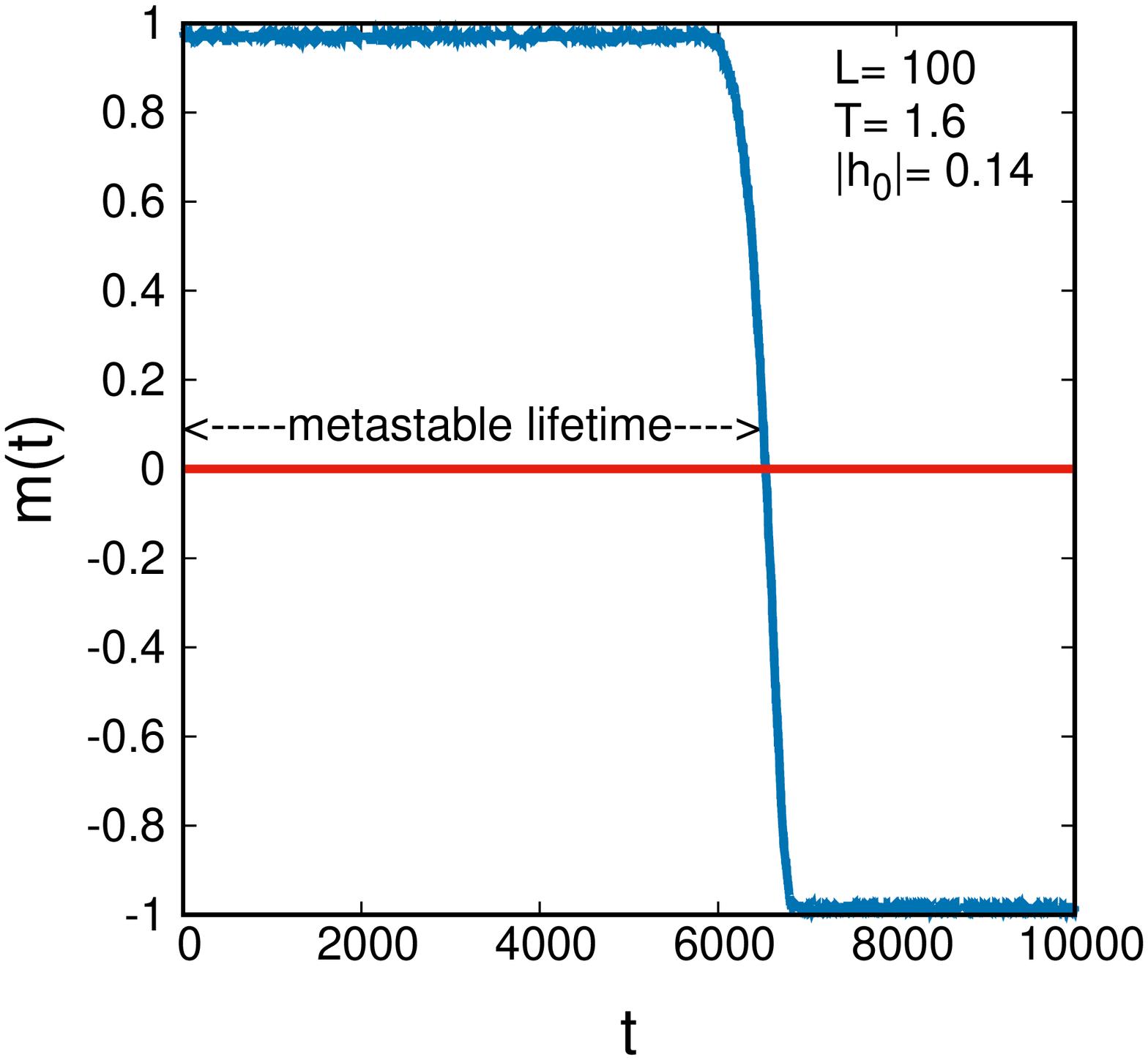}
		\subcaption{}
	\end{subfigure}
	\begin{subfigure}[b]{0.5\textwidth}
		\includegraphics[width=0.8\textwidth,angle=-90]{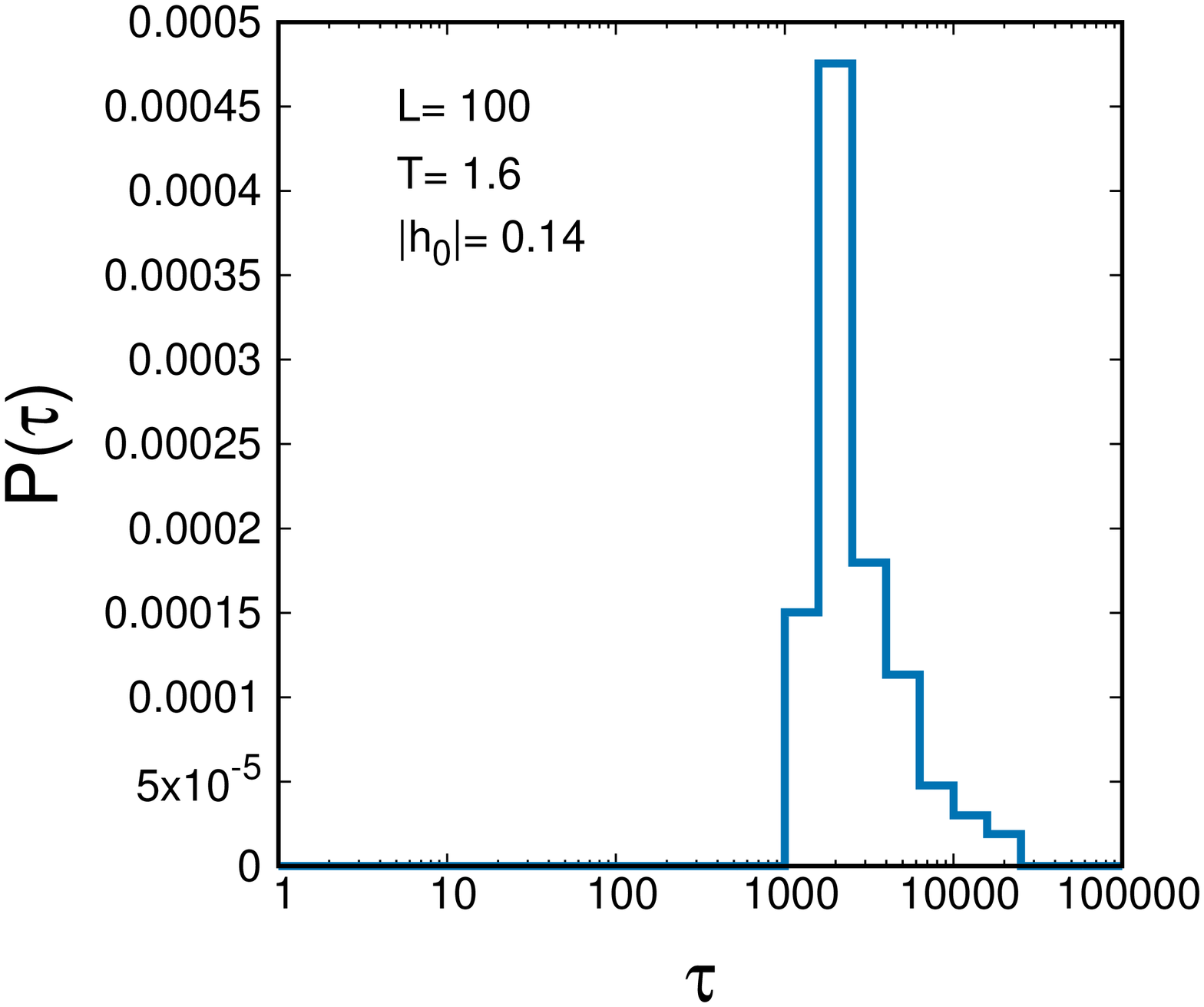}
		\subcaption{}
	\end{subfigure}
	\caption{(a) A typical decay of metastable state for lattice size $L=100$ at temperature $T=1.6(0.7T_c$ 
		in the presence of external field $h_0=-0.14$. (b) Normalised distribution of reversal times ($\tau$) for 10000 samples with the same parameters as used in (a)}
	\label{meta_state}
\end{figure}

Decay of the metastable states of matter are the well-known significant phenomena of nature. If one prepares a system in a metastable 
state, the system does not remain in this state forever, rather it decays and eventually reaches a true equilibrium state. Imagine a lattice 
with Ising spins in a small negative magnetic field $h_0=-0.14$ at a temperature ($T=1.6$) sufficiently below the 
critical temperature $T_c(=2.269..)$. We have observed that the system will keep itself in a metastable state (with positive magnetisation) even in the negative field. Then after certain period of time it will decay to a stable state with negative magnetisation. This situation is depicted in fig-\ref{meta_state}a, where the magnetisation is plotted with time. Here 
we have defined the \textit{lifetime of metastable state} or the \textit{reversal time of magnetisation} ($\tau$) as the minimum time required to achieve negative magnetisation from a completely ordered state. In the fig-\ref{meta_state}a the reversal time is 6535 MCSS. 

Fig-\ref{meta_state}a represents the decay of metastable state of a single sample. Now we have taken 10000 number of such samples and calculated the reversal 
times for each such samples. The errorbars of the data for the reversal times are
calculated from the standard deviations. Fig-\ref{meta_state}b shows normalised probability distribution of those reversal 
times obtained over 10000 number of different samples. The most probable/ peak  value of the distribution is around 2000-3000 MCSS which is certainly
different from that of the single sample (around 6500 MCSS). Due to this
reason, instead of studying the reversal time for a single, 
sample we will study the behaviour of mean/ average ($\tau_{av}$) and most probable/ peak ($\tau_{mp}$) of the reversal time in our work to show the statistical characteristics of the reversal times.

\begin{figure}[h]
\begin{center}
	\resizebox{10cm}{!}{\includegraphics[angle=-90]{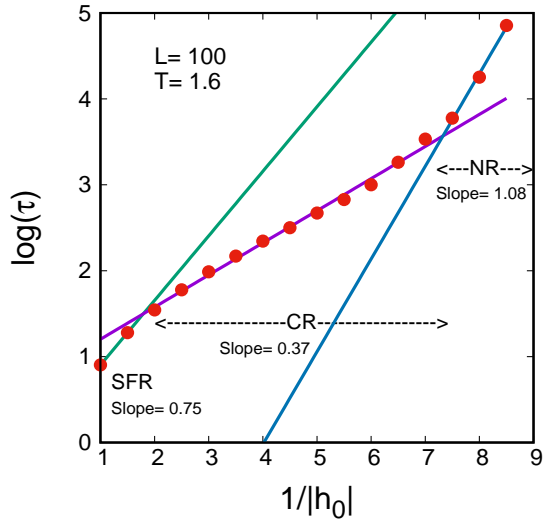}}
	\caption{log of mean reversal time with $1/h_0$ for lattice size $L=100$ at temperature $T=1.6$, 
                 where $h$ is uniform field.}
        \label{becker} 
\end{center}
\end{figure}

\begin{figure}[h!]
	\begin{subfigure}[b]{0.31\textwidth}
		\includegraphics[width=0.82\textwidth,angle=-90]{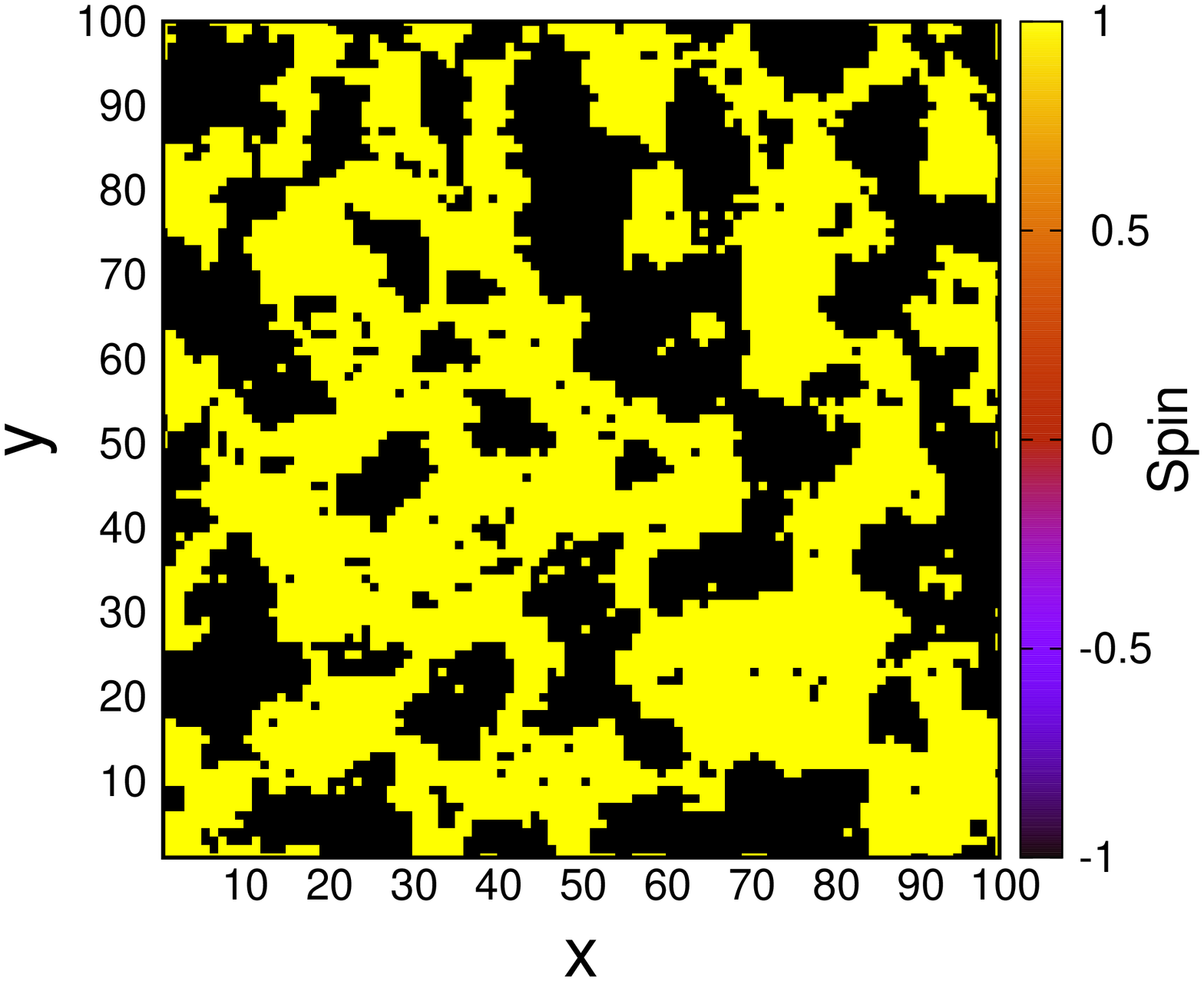}
		\subcaption{}
	\end{subfigure}
	\begin{subfigure}[b]{0.31\textwidth}
		\includegraphics[width=0.82\textwidth,angle=-90]{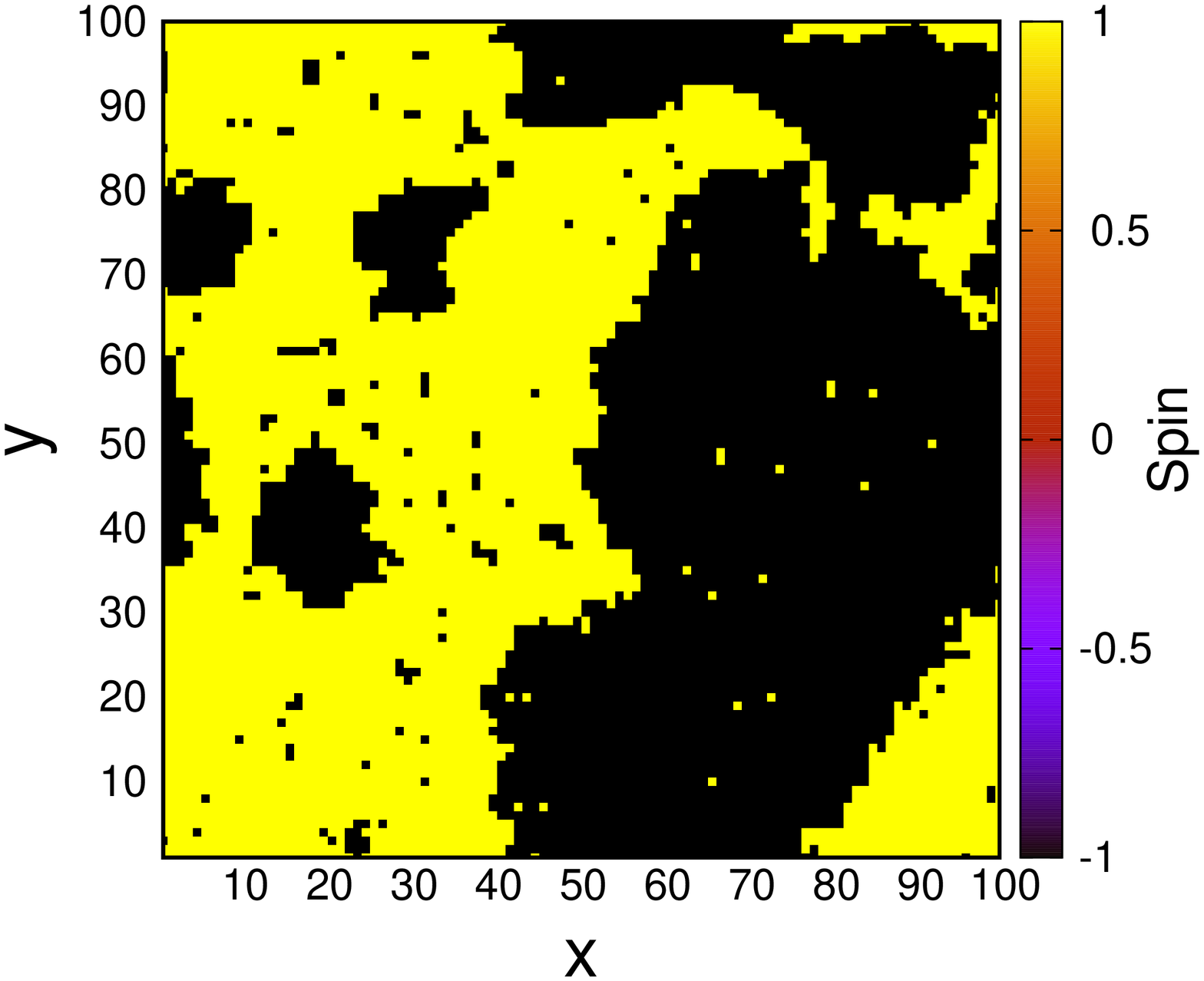}
		\subcaption{}
	\end{subfigure}
	\begin{subfigure}[b]{0.31\textwidth}
		\includegraphics[width=0.82\textwidth,angle=-90]{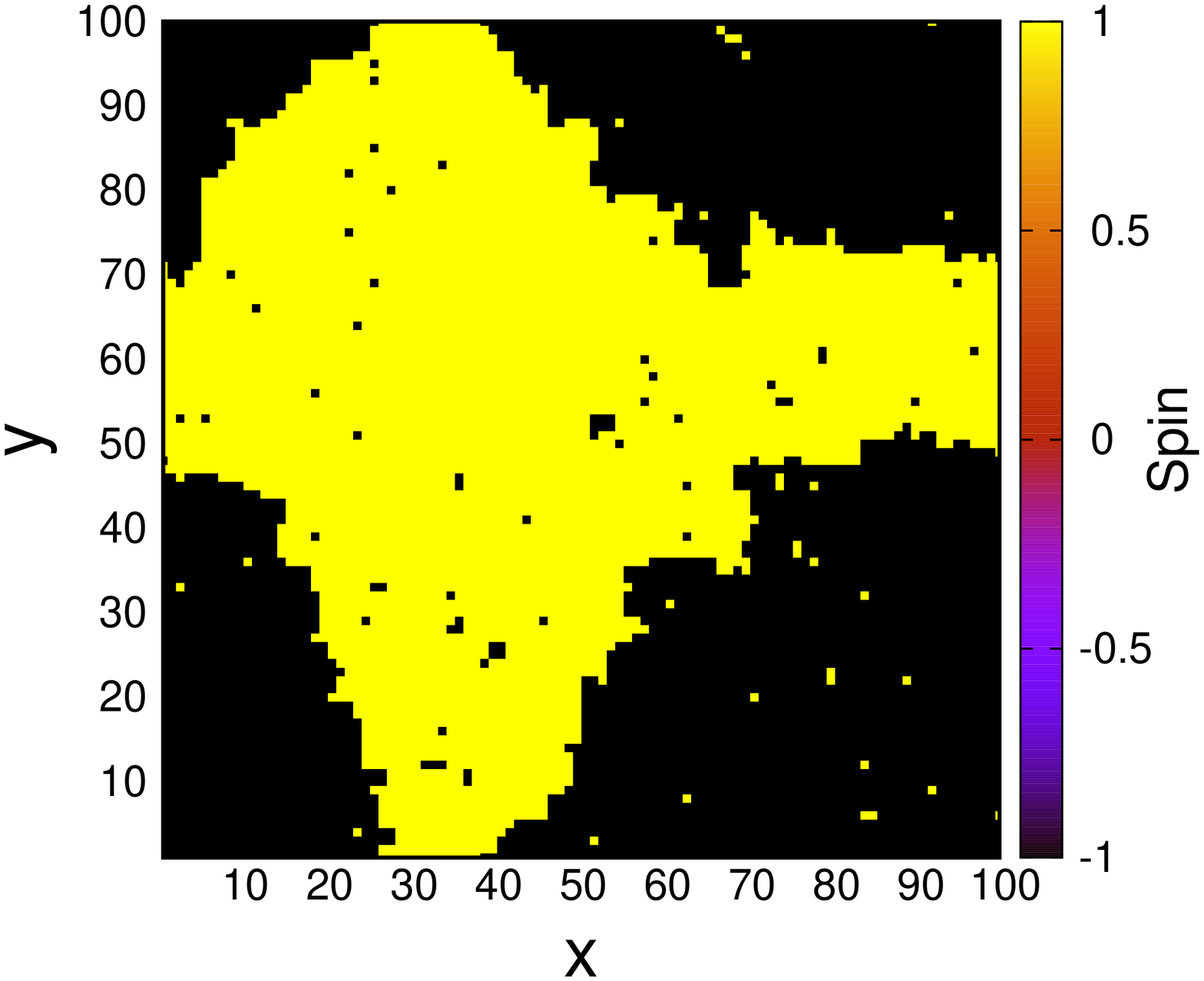}
		\subcaption{}
	\end{subfigure}
	\caption{Snapshots at the time of reversal in three regimes of log($\tau$) vs $1/|h_0|$ plot. (a) 19 MCSS for strong field regime ($1/h_0= -1.5$) (b) 235 MCSS for coalescence regime ($1/h_0= -4.5$) (c) 7815 MCSS for nucleation regime ($1/h_0= -8.0$).}
	\label{beck_snap}
\end{figure}

We have calculated mean reversal time by taking average over the reversal times obtained from 5000 ferromagnetic system and plotted logarithm of this mean reversal time with $1/h_0$ (uniform field only) for a fixed temperature. In this plot (Fig-\ref{becker}), three distinct regimes 
(strong field regime (SFR), coalescence/ multidroplet regime (CR) and nucleation/ single droplet regime (NR)) are clearly identified. The reversal times in the three regimes are separately fitted to straight lines. It is observed that 
the slope ($\simeq 0.37$) of the curve in coalescence regime is almost 1/3 ($\frac {1}{d+1}$ according to the equations 7a and 9a where $d=2$) of the slope ($\simeq 1.08$) in nucleation regime. We have taken the snapshots of spin configurations at the time of reversal in three regimes and shown in (Fig-\ref{beck_snap}). 

\begin{figure}[h!]
	\begin{subfigure}[b]{0.3\textwidth}
		\includegraphics[width=0.85\textwidth,angle=-90]{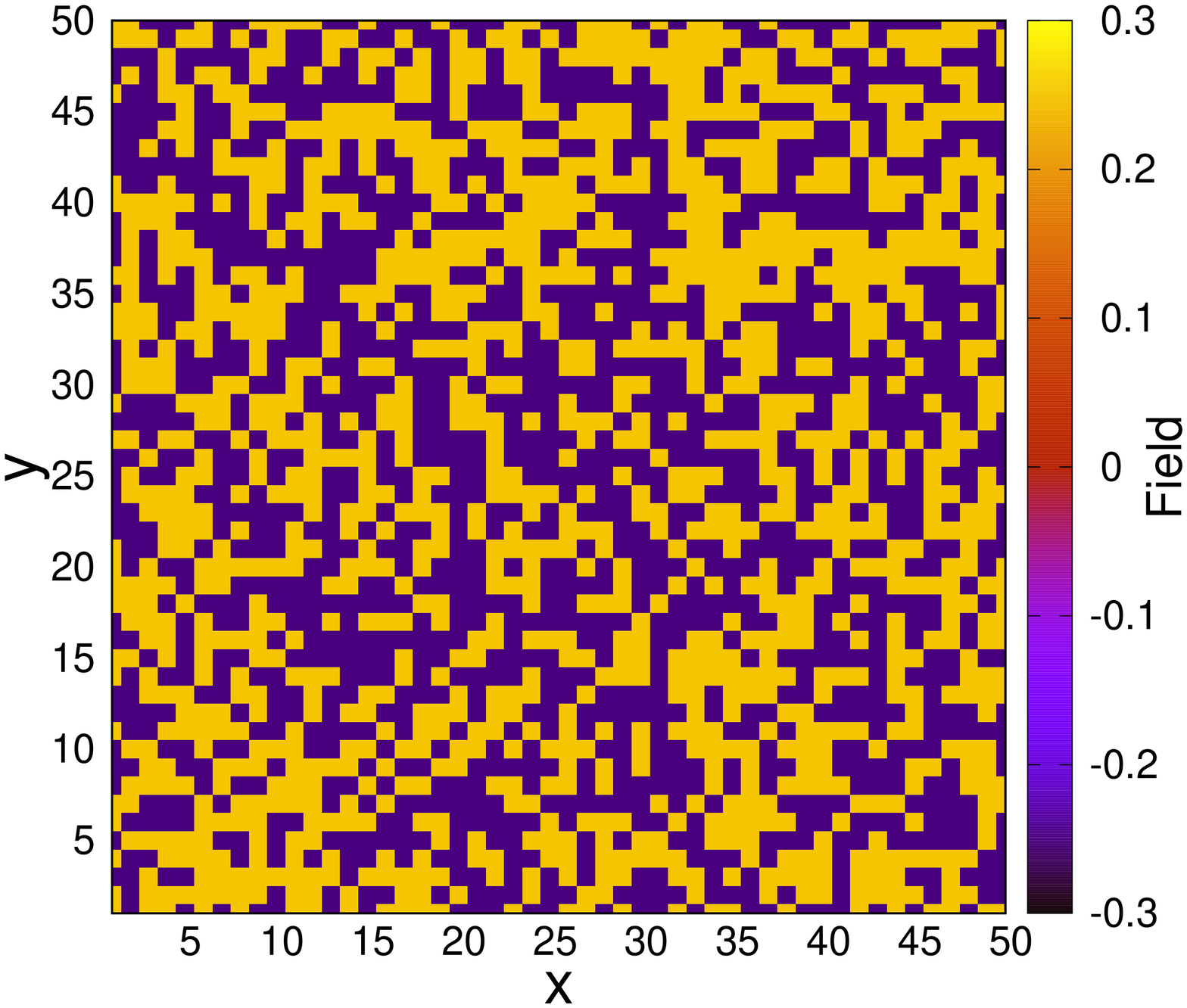}
		\subcaption{}
	\end{subfigure}
	\begin{subfigure}[b]{0.3\textwidth}
		\includegraphics[width=0.85\textwidth,angle=-90]{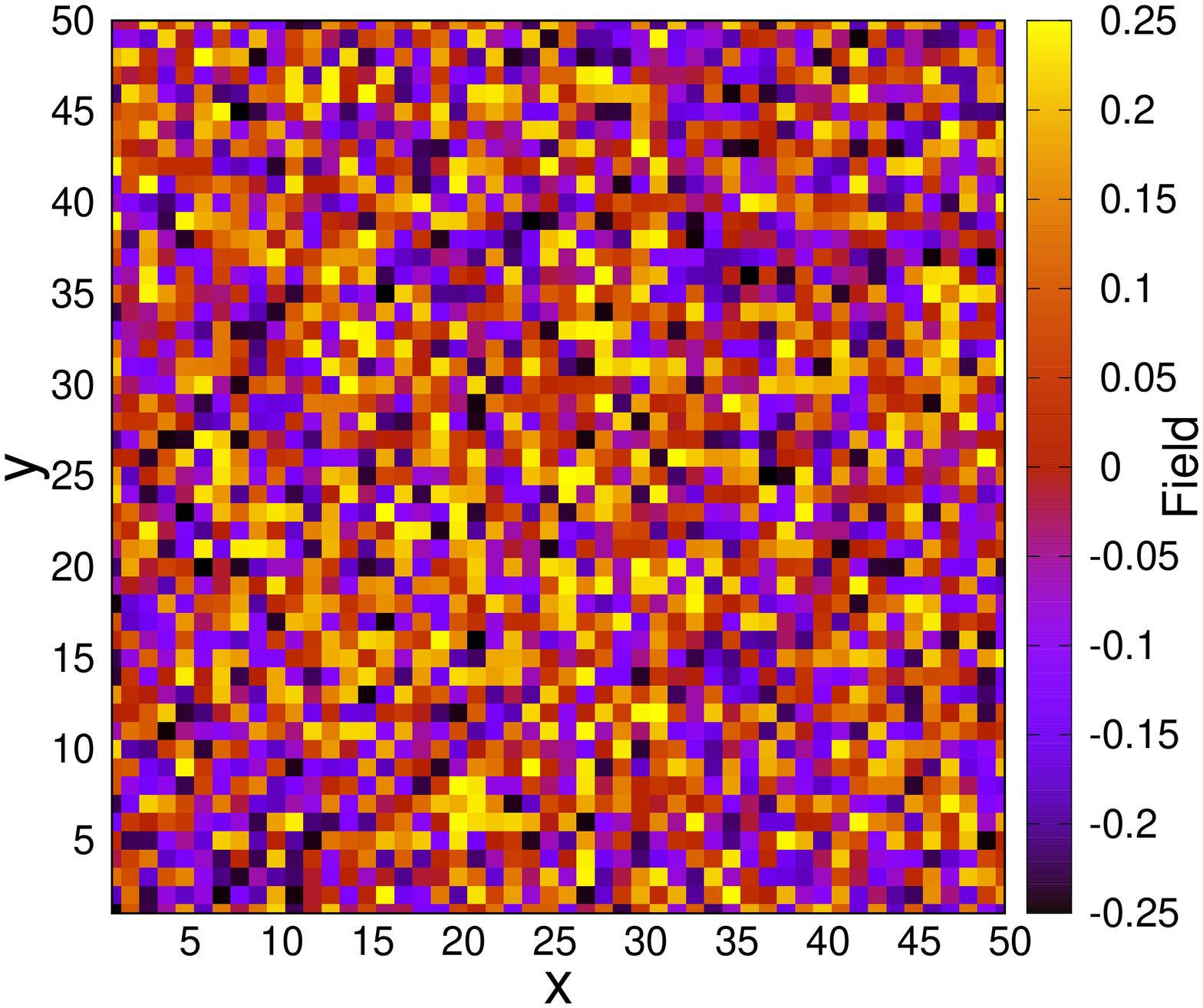}
		\subcaption{}
	\end{subfigure}
	\begin{subfigure}[b]{0.3\textwidth}
		\includegraphics[width=0.85\textwidth,angle=-90]{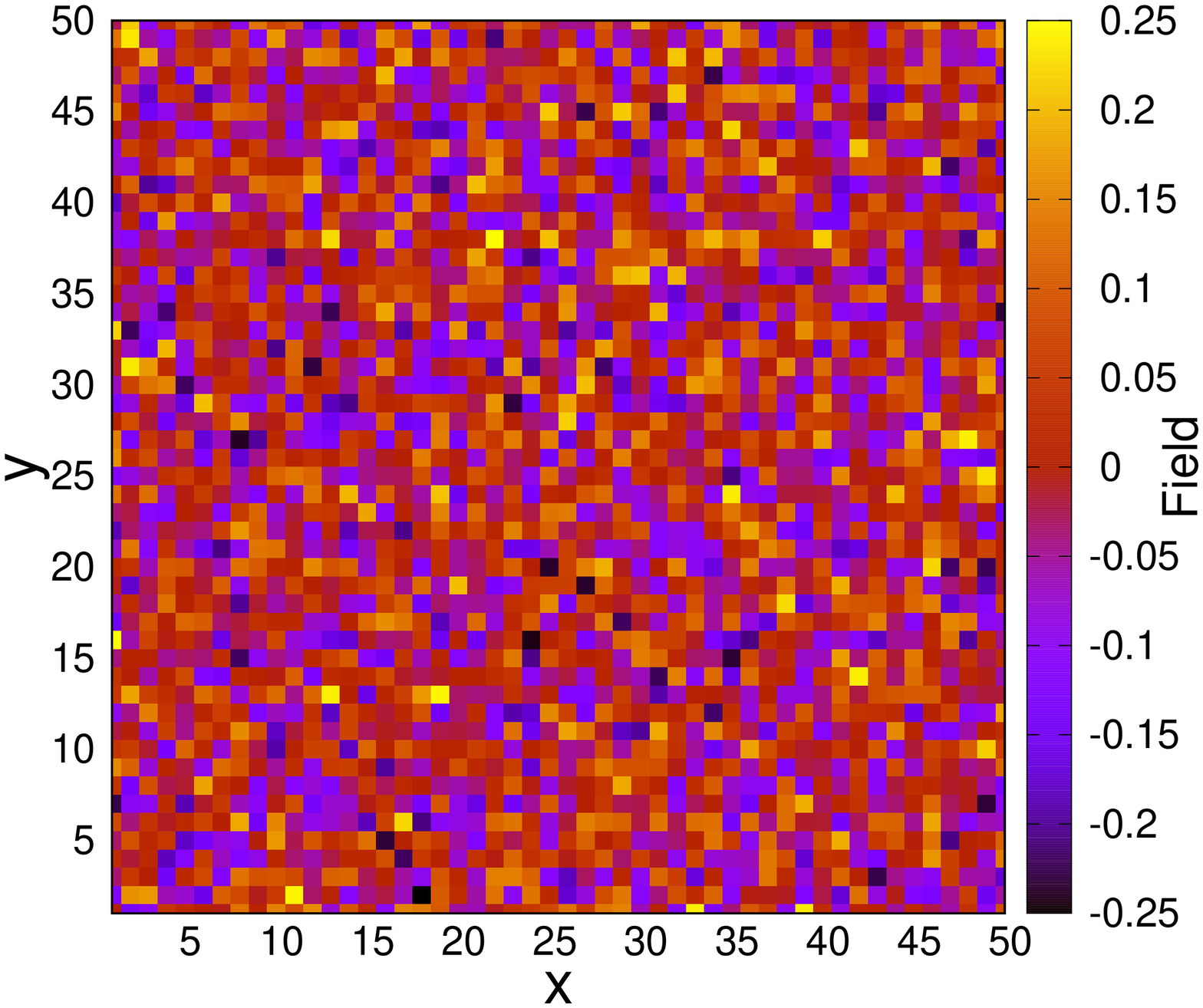}
		\subcaption{}
	\end{subfigure}
	\caption{Image plots (zoomed in from $L=300$ to $L=50$) of the three different distributions 
		of random field each of mean 0 and width $w=0.25$, (a)Bimodal random field, (b)Uniformly 
		distributed random field, (c)Gaussian random field with $\sigma=w/5$.}
	\label{field_morpho}
\end{figure}

Now we have applied a quenched (site dependent but time independent) random field ($h_r$) in addition to the already applied 
uniform magnetic field ($h_0$) to the system. So at each lattice site random fields are different, but they are fixed with time (quenched disorder). Fig-\ref{field_morpho} shows the image plots of the values of three different types of random fields ($h_r$). Where the colored-scale in the side bar 
of each plot represents different values of the random field. The Fig-\ref{field_morpho}a shows the image plot of the bimodal random field where only two colors (yellow for $h_r=0.25$ and dark blue for $h_r=-0.25$) are present in equal proportion. The Fig-\ref{field_morpho}b shows the image plot of the
uniformly distributed random field between $h_r=0.25$ and $h_r=-0.25$ where all the colors are present in equal probability. The
Fig-\ref{field_morpho}c shows the image plot of the Gaussian random field with $\sigma=w/5$ (but in our work we have used $\sigma=2w$) where colors near mean value are present in larger proportion. The Gaussian distribution of the random field has been
generated by widely used Box-Muller algorithm\cite{box}.

\begin{figure}[h!]
  \begin{subfigure}[b]{0.3\textwidth}
    \includegraphics[width=0.85\textwidth,angle=-90]{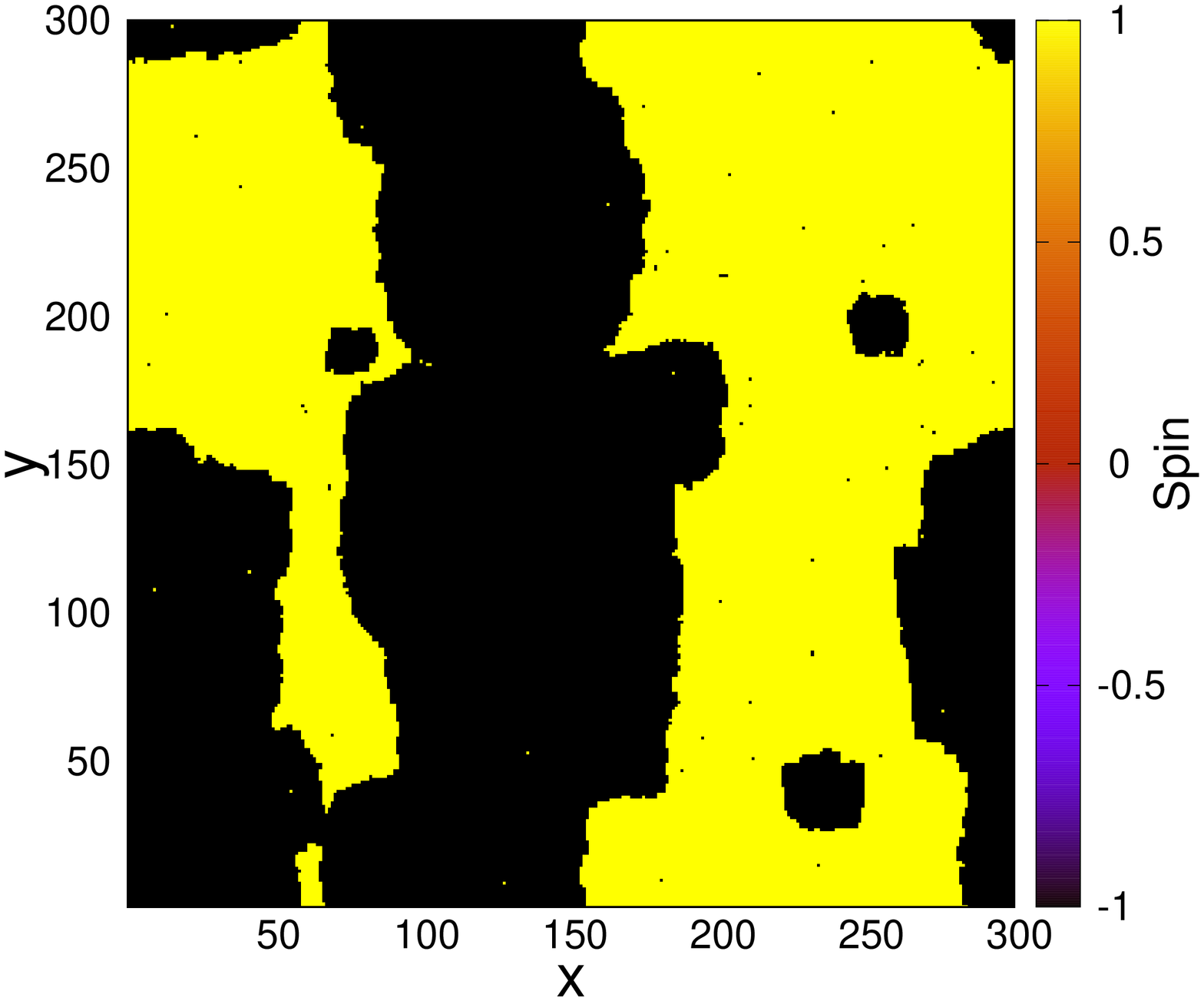}
    \subcaption{}
  \end{subfigure}
  \begin{subfigure}[b]{0.3\textwidth}
    \includegraphics[width=0.85\textwidth,angle=-90]{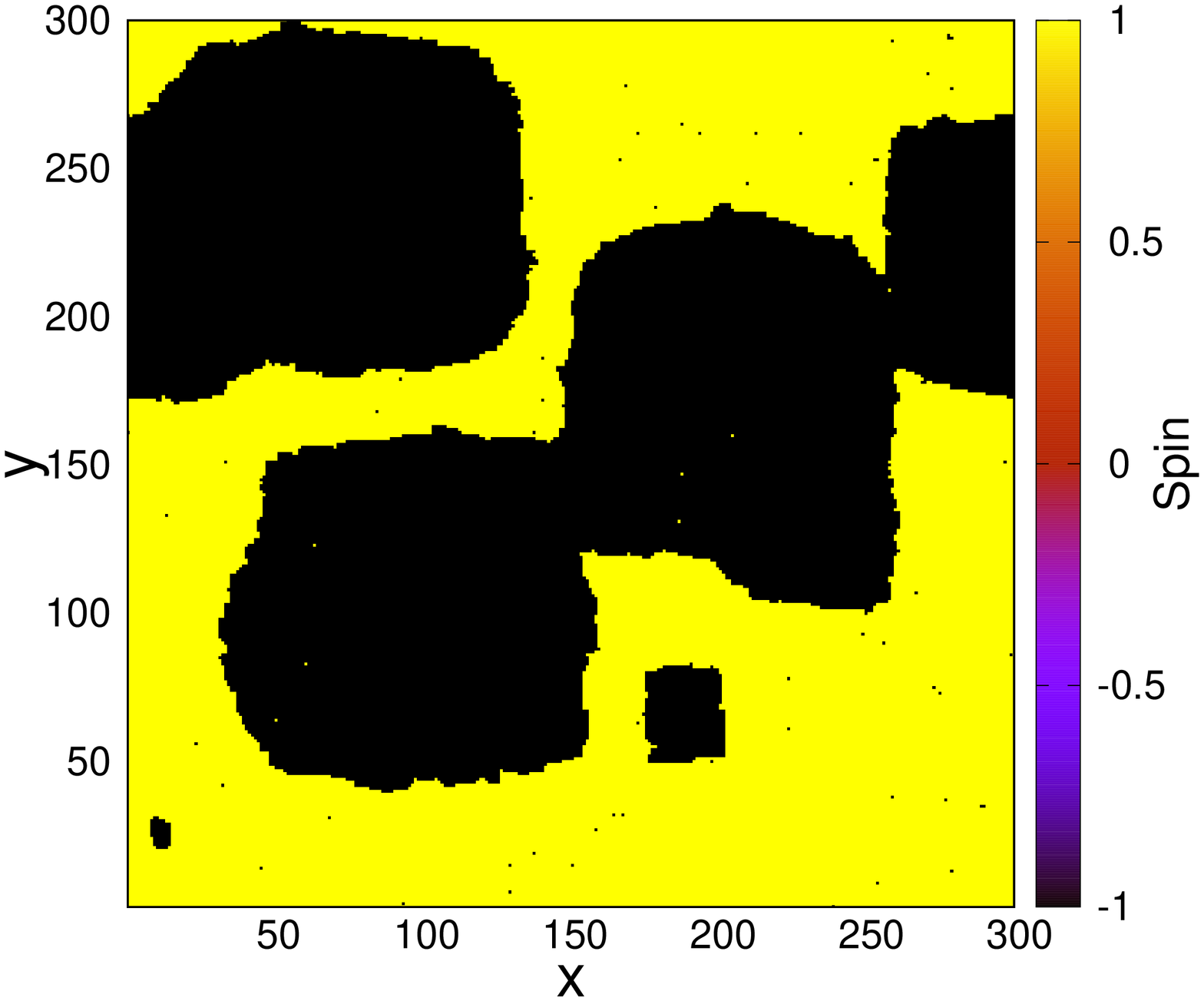}
    \subcaption{}
  \end{subfigure}
  \begin{subfigure}[b]{0.3\textwidth}
    \includegraphics[width=0.85\textwidth,angle=-90]{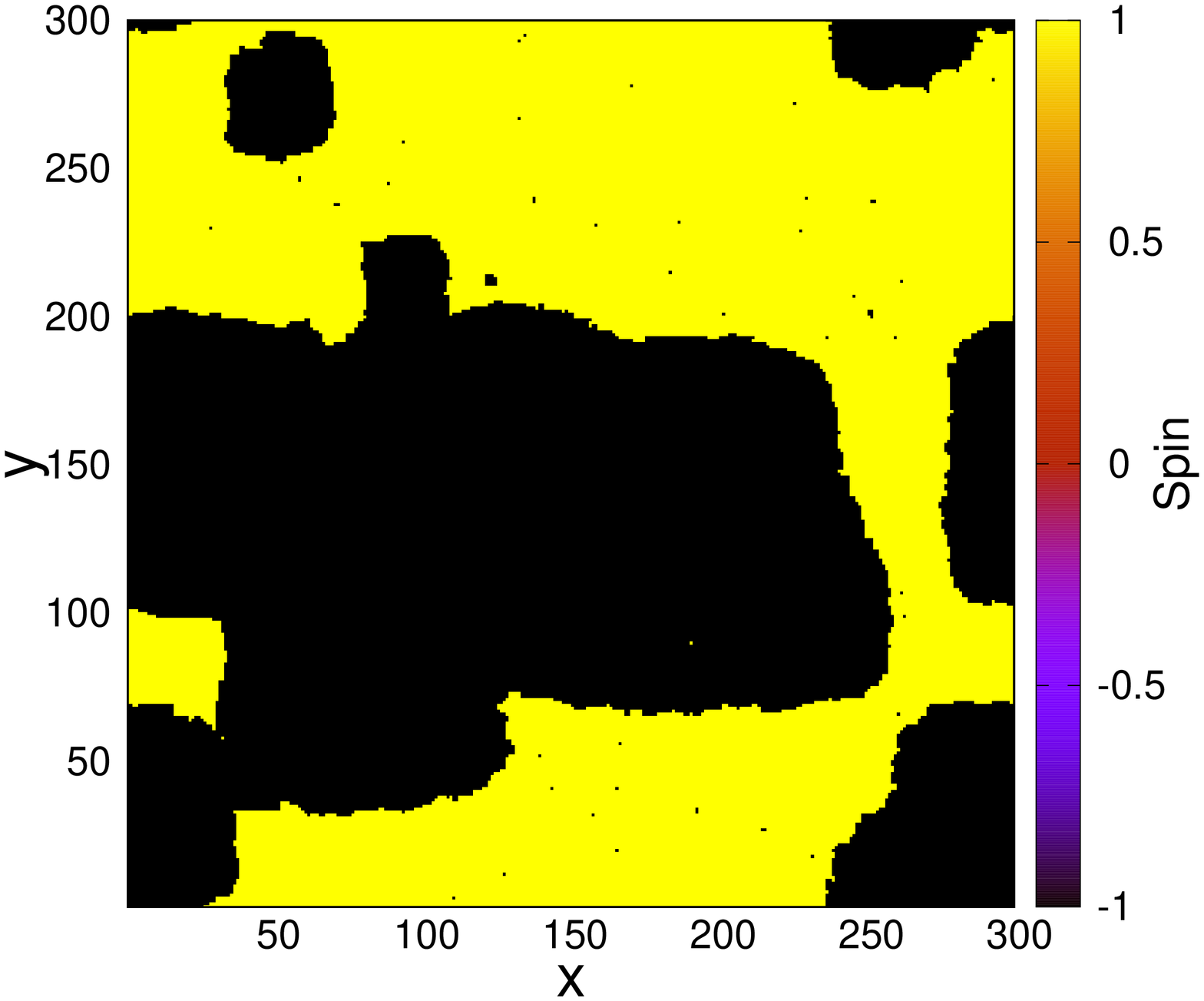}
    \subcaption{}
  \end{subfigure}
 \caption{Image plots of the values of the spins at the time of magnetisation reversal in the presence of random field ($h_i$) and $h_0=-0.5$,
	  ((a) bimodal distribution (b) uniform distribution and (c) Gaussian distribution). In each case the width of random field is set to $w=0.25$, temperature is set to $T=1.0$, 
	  lattice size is $L=300$. Snapshots are taken at (a)$t=349$ MCSS (b) $t=408$ MCSS (c)$t=418$ MCSS.}
  \label{spin_morpho}
\end{figure}

We have taken some snapshots (Fig-\ref{spin_morpho}) of the spin configurations in the presence of random field along with externally applied uniform field. Here the colors in the side bar of each plot represents the value of the spin. The yellow color represents the up (+1) spin  and the black represents the down(-1) spin. Initially we have started with perfectly ordered state. Acase for randomly distributed field thefter applying negative field small droplets of down spins are formed and at the time of magnetisation reversal almost 50 $\%$ of the lattice is covered by black color. At this time the magnetisation changes its sign. We have noticed that the reversal time of magnetisation decreases in the presence of random fields. Another interesting fact is that though each three distribution of random field is of zero mean, the spin configurations and the corresponding reversal times  are different for the three different kind of distribution of the random fields. 

\begin{figure}[h]
\begin{center}
	\resizebox{10cm}{!}{\includegraphics[angle=-90]{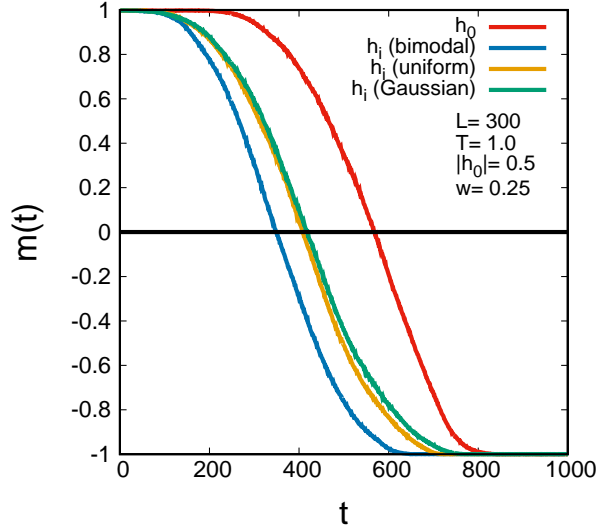}}
	\caption{Variation of magnetisation with time at temperature $T=1.0$ in presence of $h_0$ and $h_i$ (for 3 distributions of $h_r$). The value of $h_0$ is set to $h_0=-0.5$ and the width of random field 
		is set to $w=0.25$. Lattice size is $L=300$. The unit of time is MCSS (Monte Carlo step per site). }
        \label{magtime}
\end{center}
\end{figure}

To confirm the above observations,
 from the spin configurations, we have studied the evolution of the  
magnetisation $(m(t))$  with time (Fig-\ref{magtime}), in the presence of  uniform field ($h_0$) only and 
in the presence of random field $(h_i)$. The reversal times are calculated from
the decay of metastable states in all the cases. 
It is clear that the reversal time decreases in presence of any kind of random fields. Among
the three distributions, the bimodal distribution is more effective 
than the other two. For the 
Gaussian distribution, reversal time is observed to vary with its variance/width
$\sigma$. Here we have taken $\sigma= 2w$. 

\begin{figure}[h!]
	\begin{subfigure}[b]{0.5\textwidth}
		\includegraphics[width=0.8\textwidth,angle=-90]{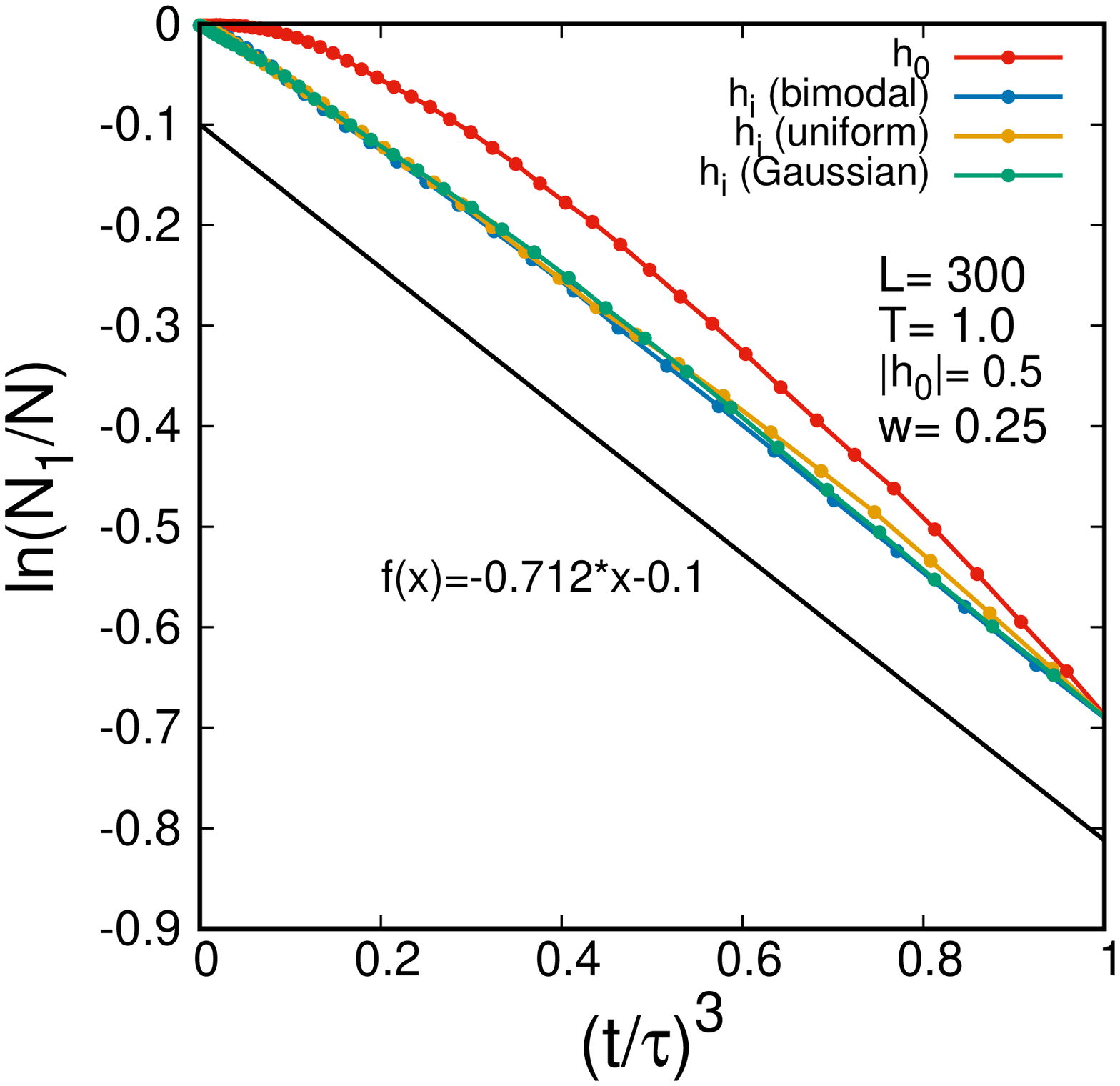}
		\subcaption{}
	\end{subfigure}
	\begin{subfigure}[b]{0.5\textwidth}
		\includegraphics[width=0.8\textwidth,angle=-90]{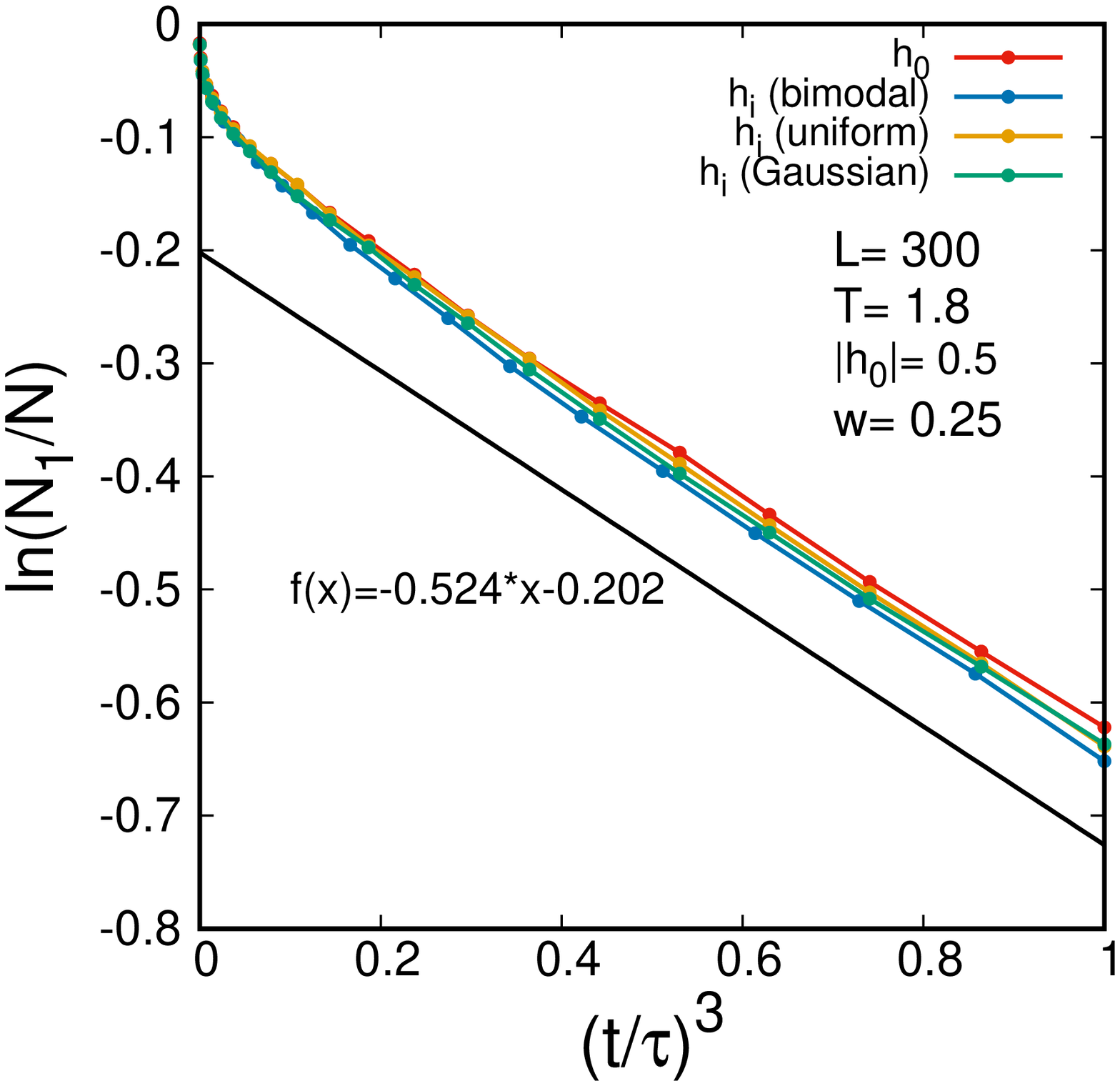}
		\subcaption{}
	\end{subfigure}
	\caption{logarithmic metastable 
	volume fraction (${{N_1} \over {N}}$, the relative
	abundance of up (+1) spins) plotted against the 
	cube of time (nondimensionalised by reversal time $\tau$) at two different temperatures (a) T=1.0 and (b) T=1.8 (0.8$T_c$). The black straight lines are the references to show that the variations of the data points are linear in the chosen scales.}
	\label{avrami}
\end{figure}

It would be interesting as well as important to know the time dependence 
of the decay of volume fraction of up (+1) spin in the presence of quenched
disorder. In the pure system, it follows Avrami's law\cite{avrami,ramos} which
states that the metastable volume fraction (relative abundance of up(+1) spins) decays with 
the third (in general, for $t^{d+1}$ for $d$ dimensional system
\cite{kolmogorov,johnson,avrami})
power of time (close to the critical temperature). We have also checked it in our
randomly queched disordered system of RFIM. Fig-\ref{avrami} shows the results of such
studies. In all three different types of distributions of random disorder, the
decay of metastable volume fractions obey the Avrami's $t^3$ law. However, the
behaviour is more prominent in the case where the temperature is closer to $T_c$
(Fig-\ref{avrami}b). It is clear from Fig-\ref{avrami} that in presence of random fields, the nature of variation of metastable volume fraction with time is almost
exponential of the third power of time, during reversal. So the system obeys the Avrami's law. But in the presence of uniform field
only, when the reversal time is larger relatively, then the exponential nature disappears. The system is found to deviate
from the Avrami's law. In a similar way, at $T=0.8 T_c$, if the value of 
the field is very low, i.e. in nucleation regime where the reversal
time is large enough, then the metastable volume fraction fails to show the
 exponential (of third power of time) nature of decay with time. This seems to
 be the possible reason of deviation
from the Avrami's law.

\begin{figure}[h!]
  \begin{subfigure}[b]{0.5\textwidth}
    \includegraphics[width=0.8\textwidth,angle=-90]{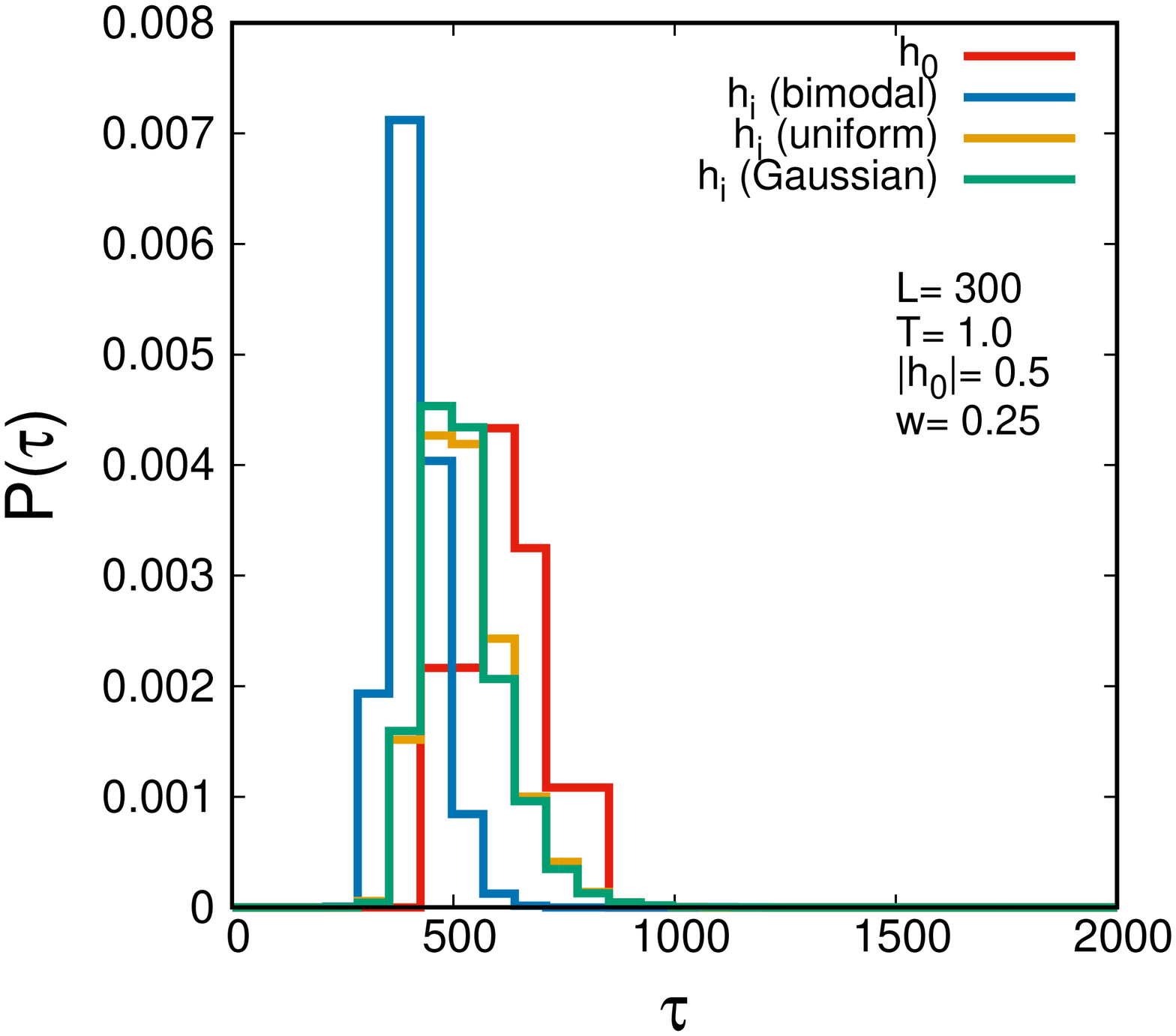}
    \subcaption{}
  \end{subfigure}
  \begin{subfigure}[b]{0.5\textwidth}
    \includegraphics[width=0.8\textwidth,angle=-90]{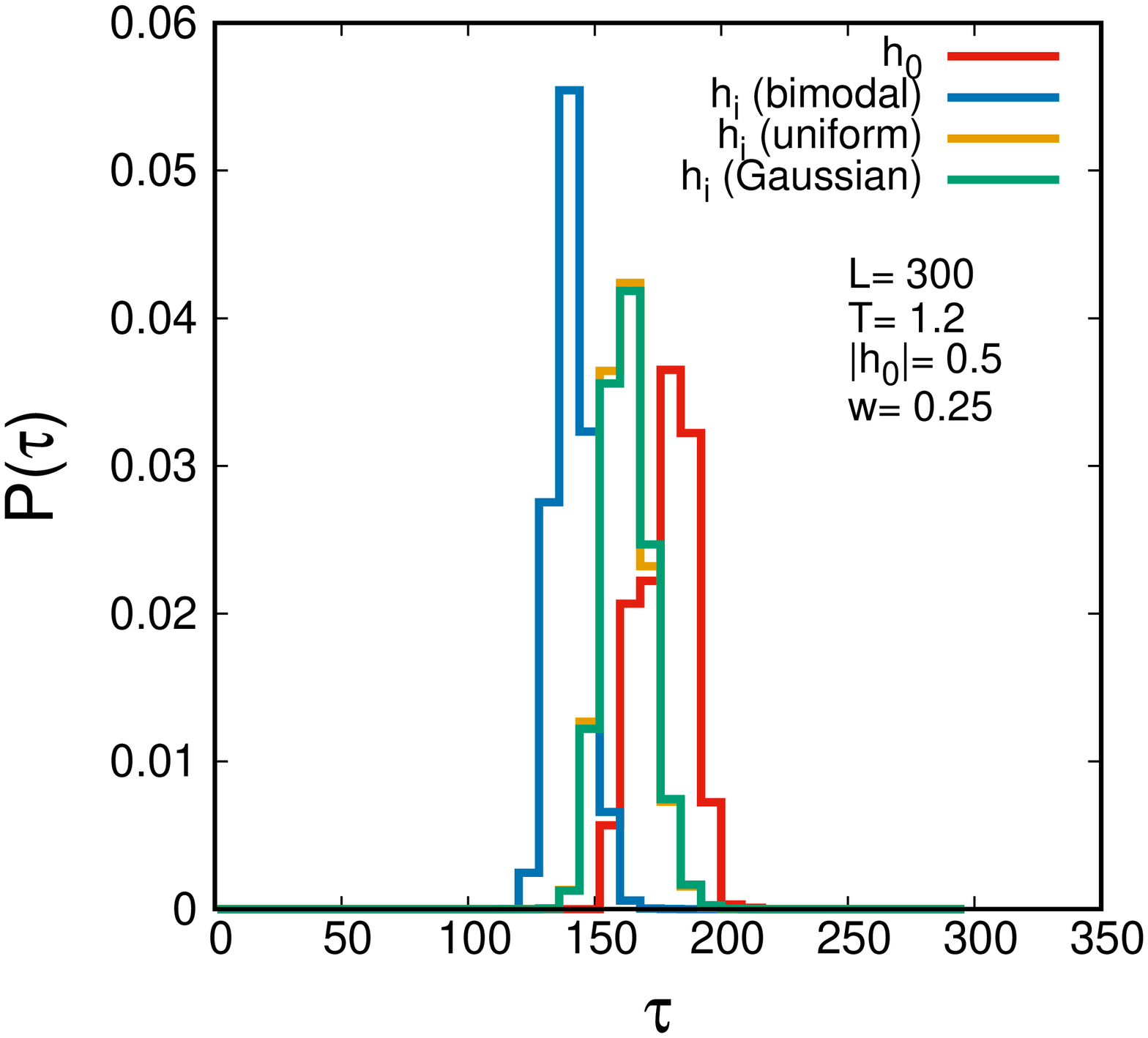}
    \subcaption{}
  \end{subfigure}
  \begin{subfigure}[b]{0.5\textwidth}
    \includegraphics[width=0.8\textwidth,angle=-90]{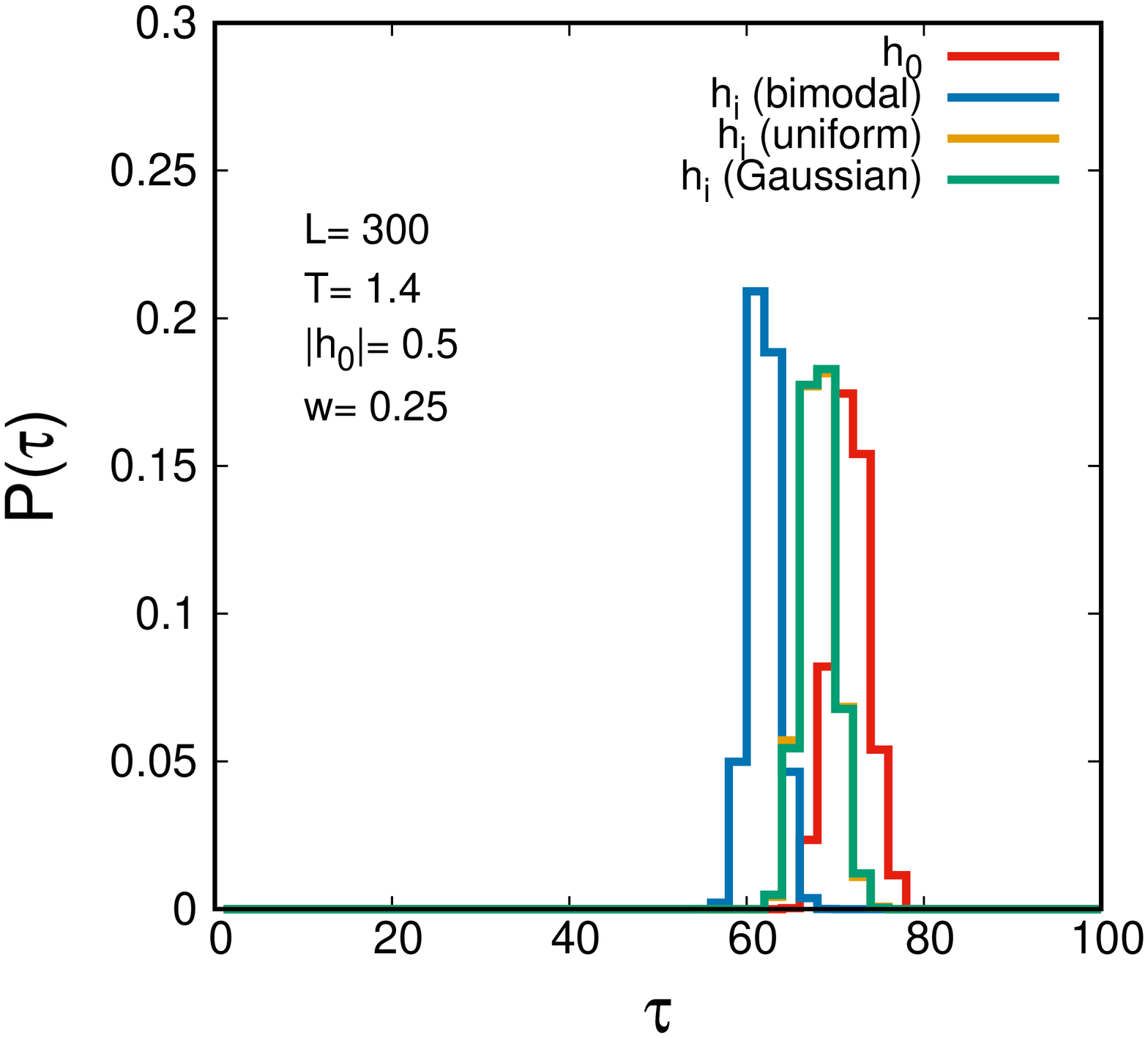}
    \subcaption{}
  \end{subfigure}
  \begin{subfigure}[b]{0.5\textwidth}
    \includegraphics[width=0.8\textwidth,angle=-90]{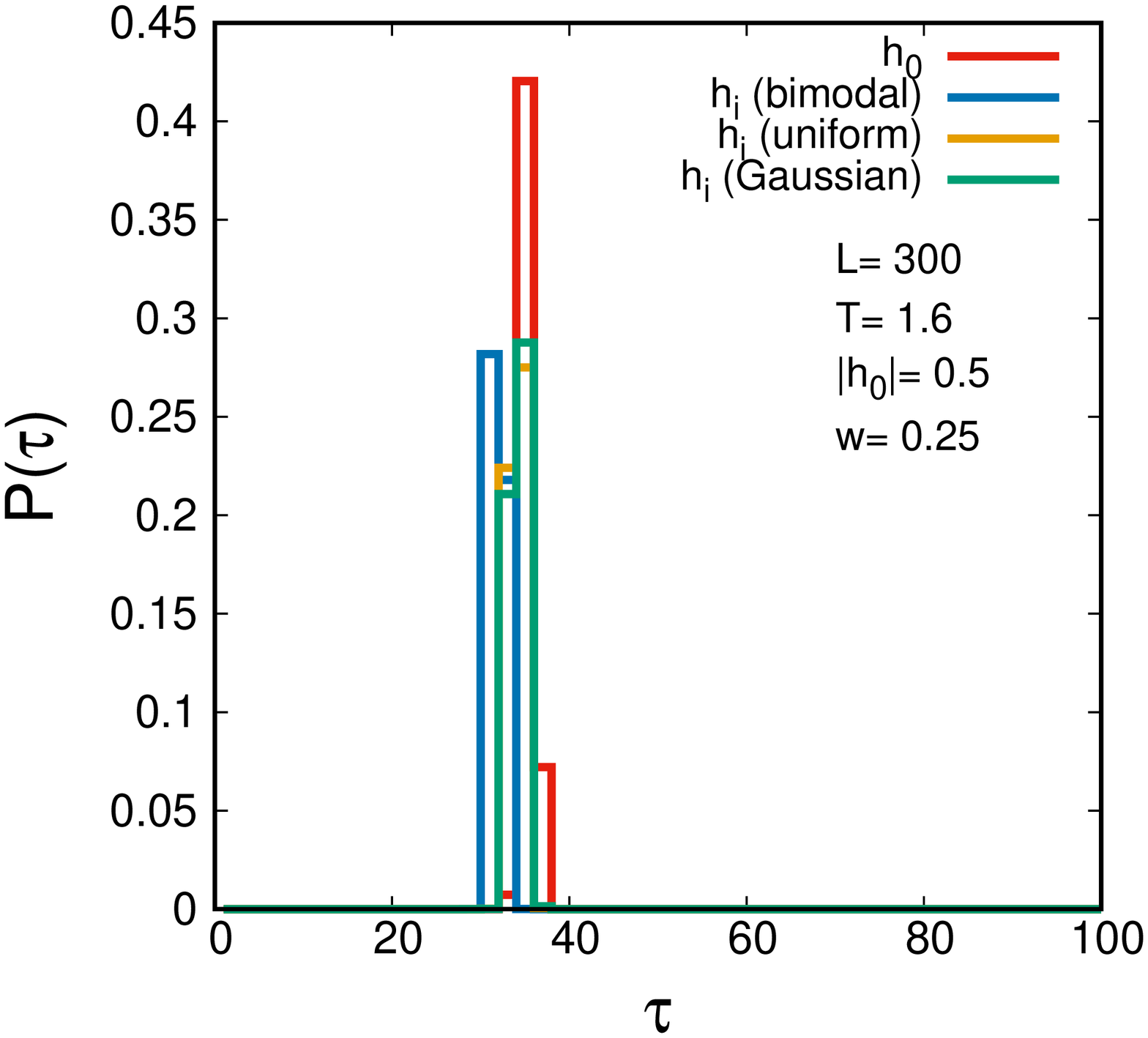}
    \subcaption{}
  \end{subfigure}
\caption{Normalised distribution of reversal times for 10000 different samples at four different temperatures
        (a)$T=1.0$ (b)$T=1.2$ (c)$T=1.4$ (d)$T=1.6$ in the presence of $h_0$ and $h_i$ (for 3 distributions of $h_r$). 
	Lattice size is $L=300$, the value of uniform field is $h_0=-0.5$ 
	and the width of each random field is $w=0.25$.}
\label{dist_temp}
\end{figure}

Fig-\ref{dist_temp} depicts normalised probability distribution of the reversal times obtained 
from 10000 different samples. The samples differ by the seeds of the random number generator used for the decision making of the flip of spin in Metropolis algorithm. Here also, we observed that the most probable reversal time decreases in the 
presence of random field. We have plotted these distributions for the four different temperatures of the system. 
It is observed that the distributions get closer to each other and also the variance decreases 
 with the 
increase in temperature. Actually at higher temperature regime, thermal noise comes into play and dominates over the 
distributions of the field. 
That is why the reversal times are almost same for any kind of the 
distributions of the field. 

\begin{figure}[h!]
  \begin{subfigure}[b]{0.5\textwidth}
    \includegraphics[width=0.8\textwidth,angle=-90]{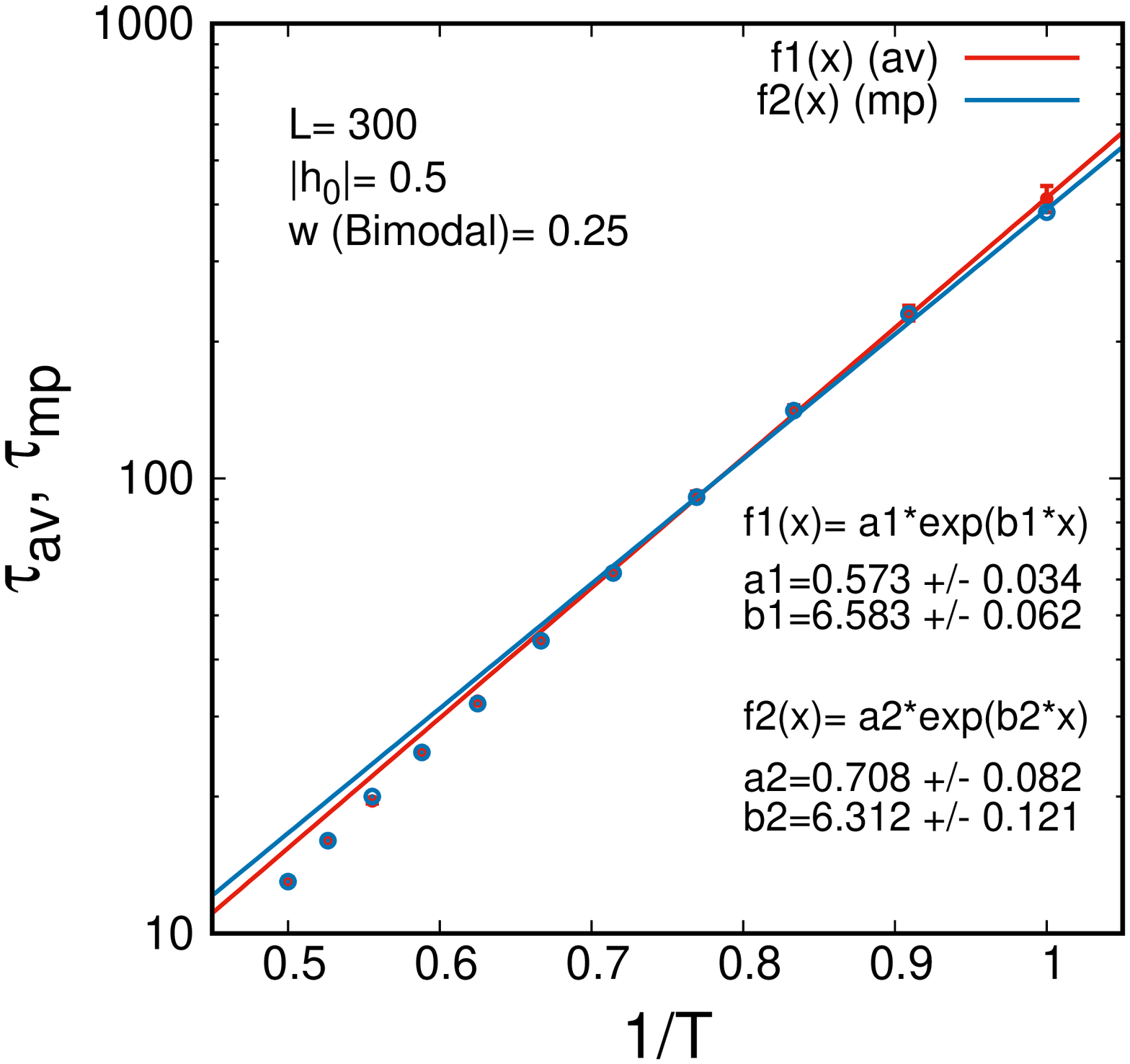}
    \subcaption{}
  \end{subfigure}
  \begin{subfigure}[b]{0.5\textwidth}
    \includegraphics[width=0.8\textwidth,angle=-90]{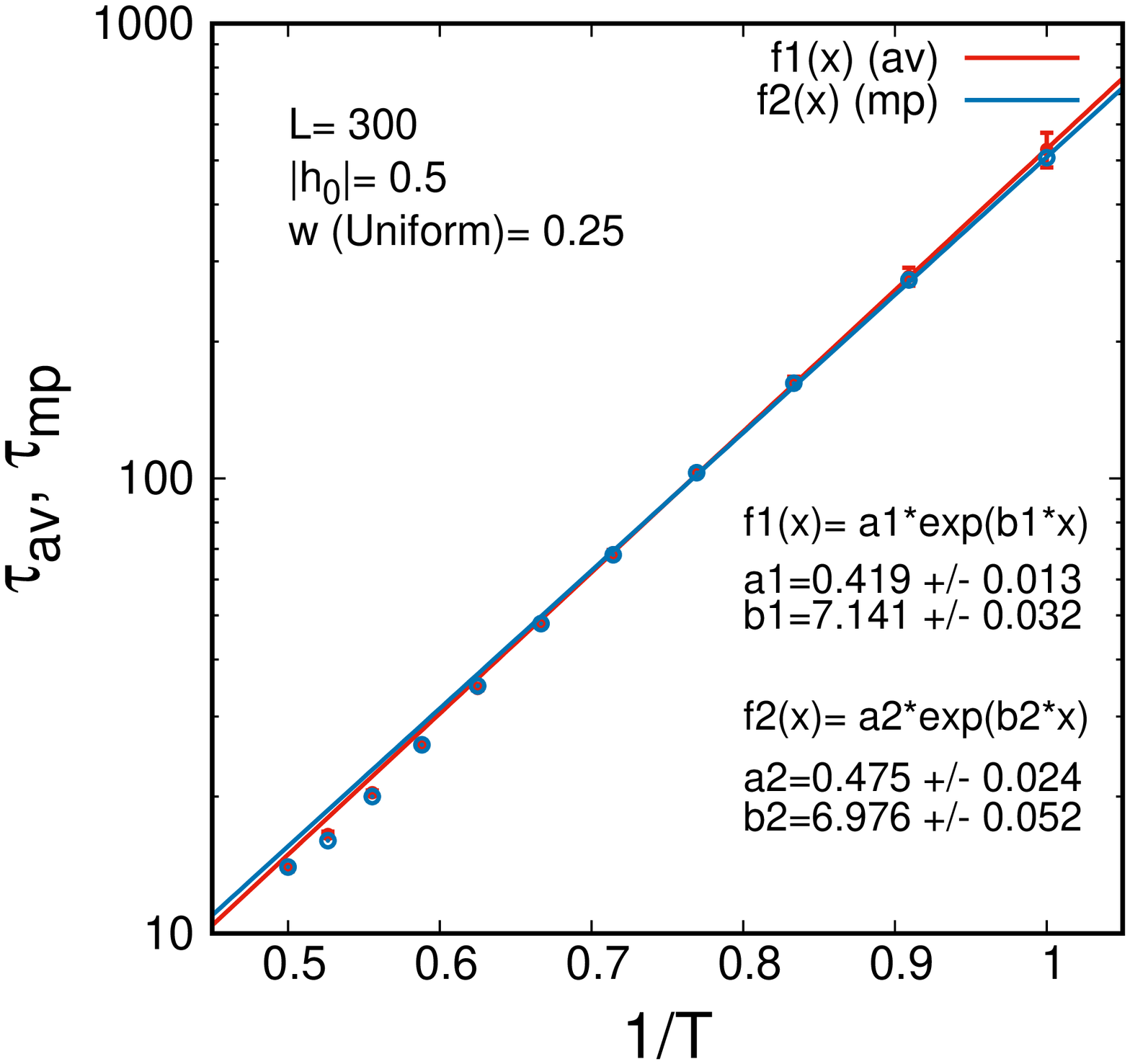}
    \subcaption{}
  \end{subfigure}
  \begin{center}
  \begin{subfigure}[b]{0.5\textwidth}
    \includegraphics[width=0.8\textwidth,angle=-90]{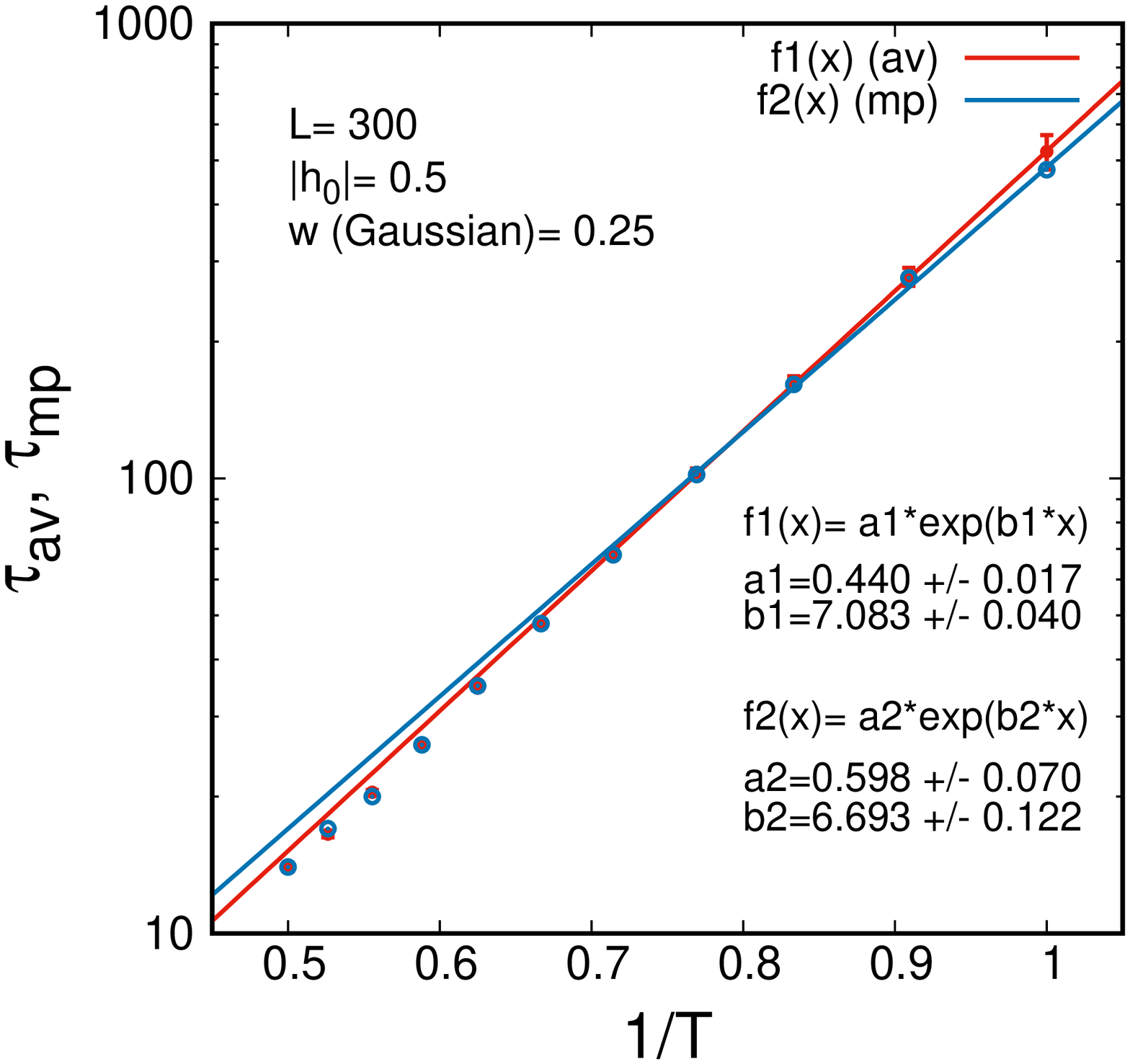}
    \subcaption{}
  \end{subfigure}
\end{center}
\caption{Variation of mean reversal time ($\tau_{av}$) and most probable reversal time ($\tau_{mp}$) with the inverse of temperature
	in presence of the random field $h_i$ ((a) bimodal (b) uniform and  
	(c) Gaussian distribution). In each case, uniform field is $h_0=-0.5$ 
	and the width of the random field is $w=0.25$, lattice size is $L=300$.}
\label{temp_var}
\end{figure}

Variation of the reversal times with temperature has been studied in presence of three different distributions of random fields having a common value of the width $w=0.25$.
Fig-\ref{temp_var} depicts that as the temperature decreases, both the mean and the most probable reversal time increase exponentially
($exp(1/{K_BT})$). The reversal times are plotted in logscale and fitted to straight line. Variance of the reversal times increases also with the decrease in temperature. 

\begin{figure}[h!]
  \begin{subfigure}[b]{0.5\textwidth}
    \includegraphics[width=0.8\textwidth,angle=-90]{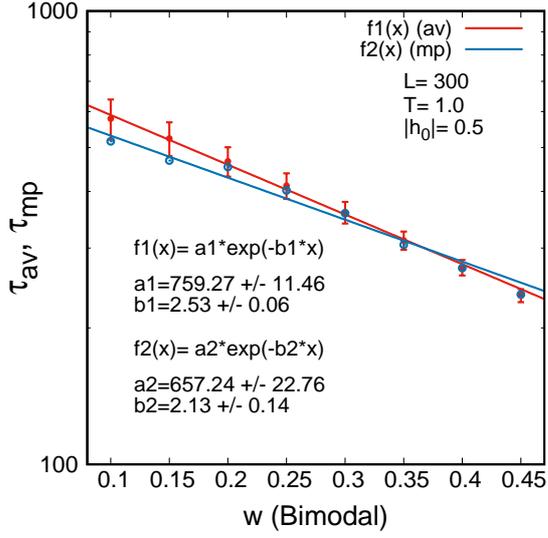}
    \subcaption{}
  \end{subfigure}
  \begin{subfigure}[b]{0.5\textwidth}
    \includegraphics[width=0.8\textwidth,angle=-90]{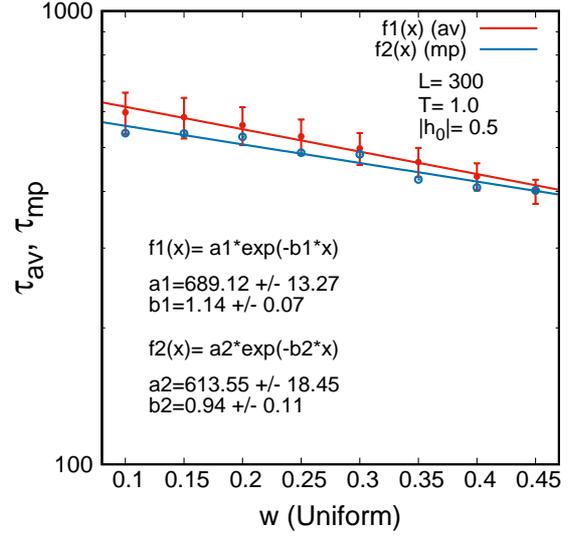}
    \subcaption{}
  \end{subfigure}
  \begin{center}
  \begin{subfigure}[b]{0.5\textwidth}
    \includegraphics[width=0.8\textwidth,angle=-90]{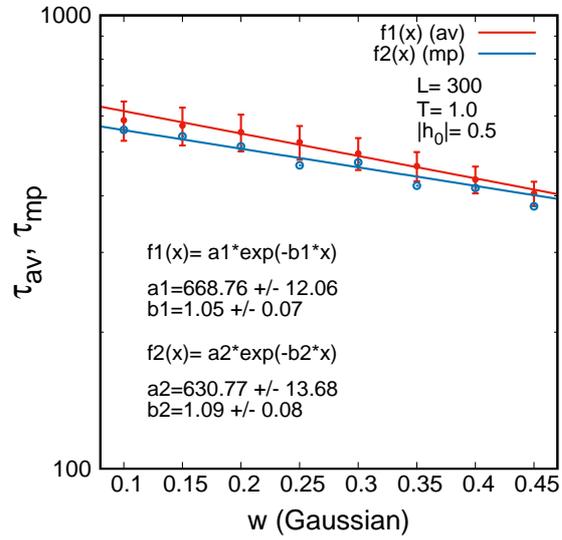}
    \subcaption{}
  \end{subfigure}
\end{center}
\caption{Variation of mean reversal time ($\tau_{av}$) and most probable reversal time ($\tau_{mp}$) with the width of random field $h_i$ (for (a) bimodal (b) uniform and  
	(c) Gaussian distribution). In each case, uniform field is $h_0=-0.5$ 
	and the temperature is $T=1.0$, lattice size is $L=300$.}
\label{width_var}
\end{figure}

In (Fig-\ref{width_var})we have studied the changes in reversal times and also their variance with the strength or width of random field disorder. Reversal times decrease exponentially with the increase in $w$. So a distribution of random field having larger width $w$ will be more effective in reducing reversal time with respect to a distribution having a smaller width. And also the variance of the reversal times decrease with the increase in $w$. So one can tune the reversal time of magnetisation by varying temperature or width of random field. Here, in a qualitative sense, the random field disorder ($w$) is equivalently playing the role
\cite{skma} of temperature ($T$). The random fields are acting as quenched disorder.
The disorder is helping to decay the metastable states. This random field acts
as an disordering field which enhances the equivalent role of the thermal fluctuations. So, in the presence
of random field the reversal occurs earlier than that compared to the absence of
random fields.

\begin{figure}[h]
	\begin{center}
		\resizebox{10cm}{!}{\includegraphics[angle=-90]{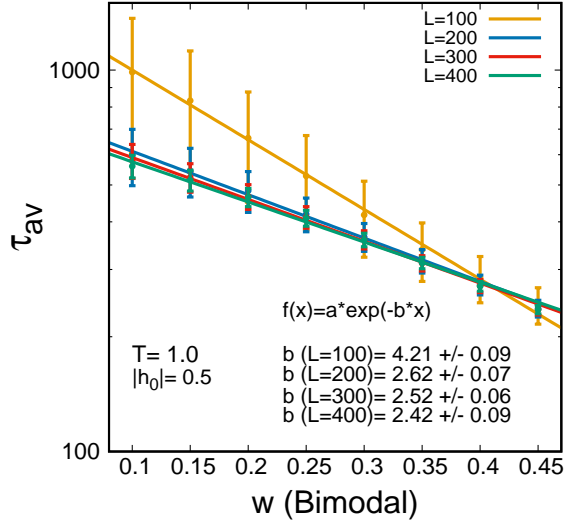}}
		\caption{Variation of mean reversal time ($\tau_{av}$) with the width of the random field $h_i$ (bimodal distribution) with different size of the lattice ($L= 100, 200, 300, 400$). In each case, uniform field is $h_0=-0.5$.}
		\label{finite_size} 
	\end{center}
\end{figure}

Any many body statistical study suffers from finite size effect. Keeping this in mind,
we have studied the reversal times as function of the width of disorder
(bimodal distribution of quenched disorder) for four different system sizes 
(namely $L=100, 200, 300$ and 400). The results are shown in Fig-\ref{finite_size}. Which shows that the most of the results obtained for $L=300$ here, do not deviate significantly from those for $L=400$. The size ($L=300$) considered here in our
present study is free from any finite size effect.

\begin{figure}[h!]
\begin{center}
	\resizebox{10cm}{!}{\includegraphics[angle=-90]{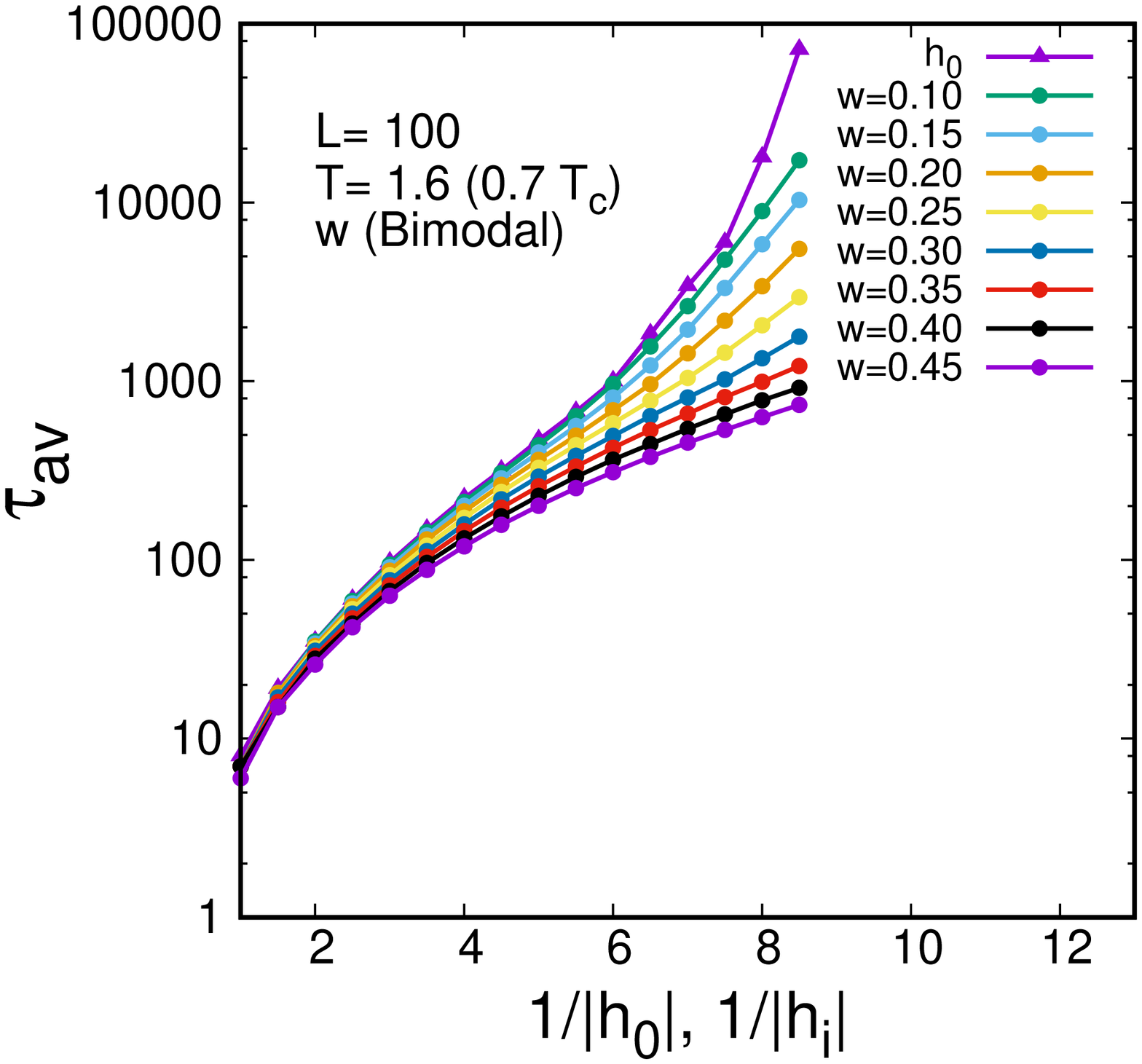}}
	\caption{Variation of the mean reversal time (in logarithmic scale)  with the inverse of field for lattice size $L=100$ 
		at temperature $T=1.6 (0.7 T_c)$ in the presence of $h_0$ (purple curve 
		with triangular points) and $h_i$ with bimodal distribution of different widths $w$ (all other 
		curves with circular points).}
\label{becker_uni}
\end{center}
\end{figure}

Since we are dealing with a two dimensional($d=2$) Ising ferromagnet, to study Becker-D\"{o}ring analysis we have studied (Fig-\ref{becker_uni}) the variations of average reversal time '$\tau$' (plotted in log scale) with the 'inverse of applied 
magnetic field' for system size $L=100$ (for affordable computational time) at temperature $T=1.6(0.7 T_c)$. In 
the presence of uniform field only, the strong field regime (SFR), the coalescence regime (CR) and 
the nucleation regime (NR)are clearly identified. Now when we apply random field into the system then the strong field and the coalescence regimes are not affected but the weak field regime i.e. the nucleation regime becomes unclear slowly. Because in nucleation regime, the field is very weak. So for a distribution of random field having stronger strength or larger width ($w > |h_0|$) which exceeds the value of uniform field ($h_0$) makes the field ($h_i$) effectively stronger than ($h_0$). Say the system is in nucleation regime with $h_0= -0.125$ or $\frac{1}{|h_0|} = 8$. Now if we apply a bimodal random field ($h_r$) having width $w= 0.3$ i.e. either $h_i= -0.125-0.3 = -0.425$ or $h_i= -0.125+0.3 = 0.175$. Now this 50$\%$ net positive field is unable to flip the spin. So the effective strength of field would be $h_i= -0.425$ which is stronger compared to $h_0= -0.125$ but affect only 50$\%$ sites approximately. That is why upto certain value of $w > |h_0|$ the system still remains in nucleation regime, showing the reversal via the growth of single droplet. But after that critical value of $w > |h_0|$, the nucleation regime disappears completely and the system stays in multidroplet regime. This does not mean that the Becker-Doring theory fails in this case. In this case, we still find the reversal via the coalescence of multiple droplets. 

\begin{figure}[h!]
  \begin{subfigure}[b]{0.5\textwidth}
    \includegraphics[width=0.7\textwidth,angle=-90]{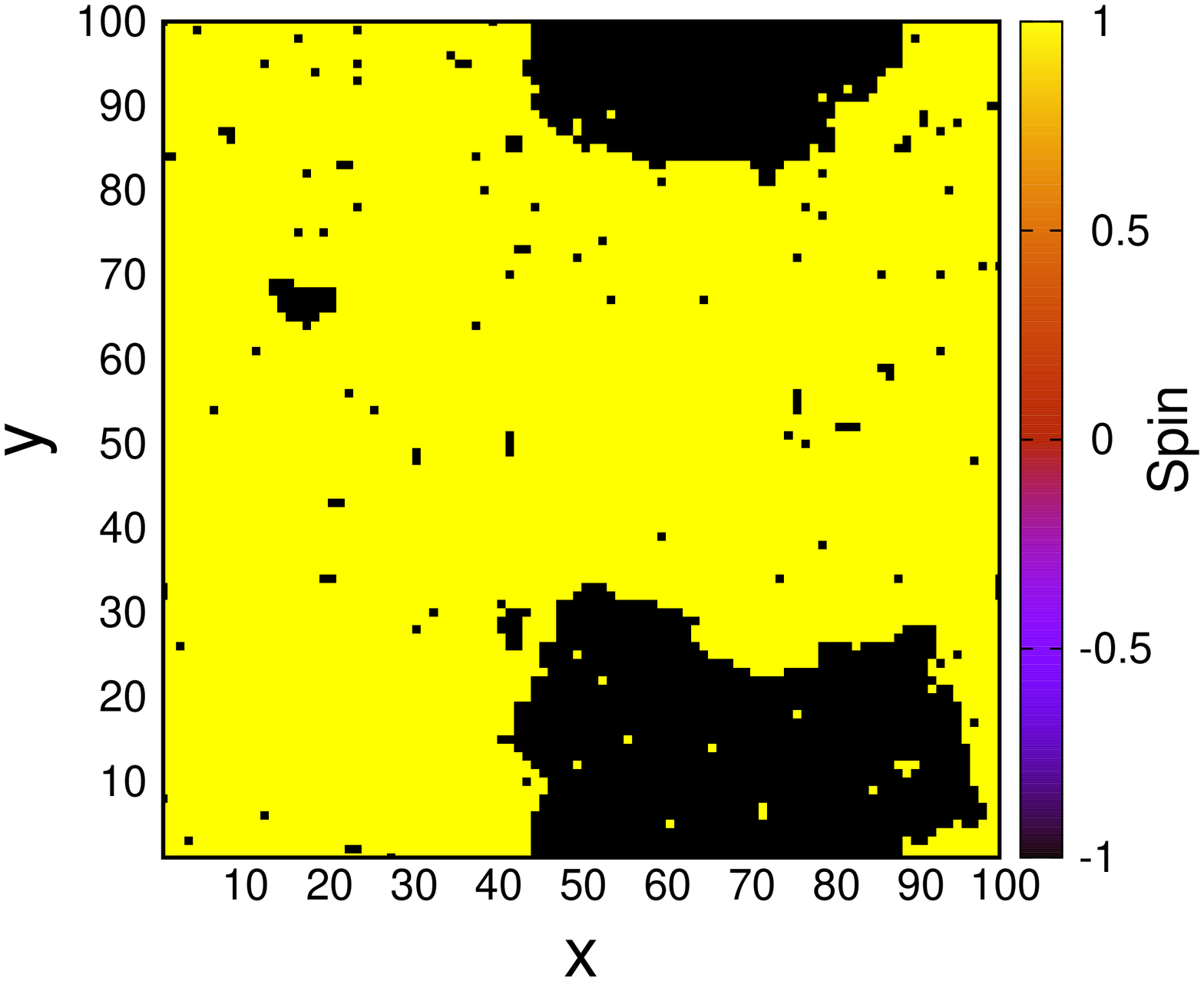}
    \subcaption{}
  \end{subfigure}
  \begin{subfigure}[b]{0.5\textwidth}
    \includegraphics[width=0.7\textwidth,angle=-90]{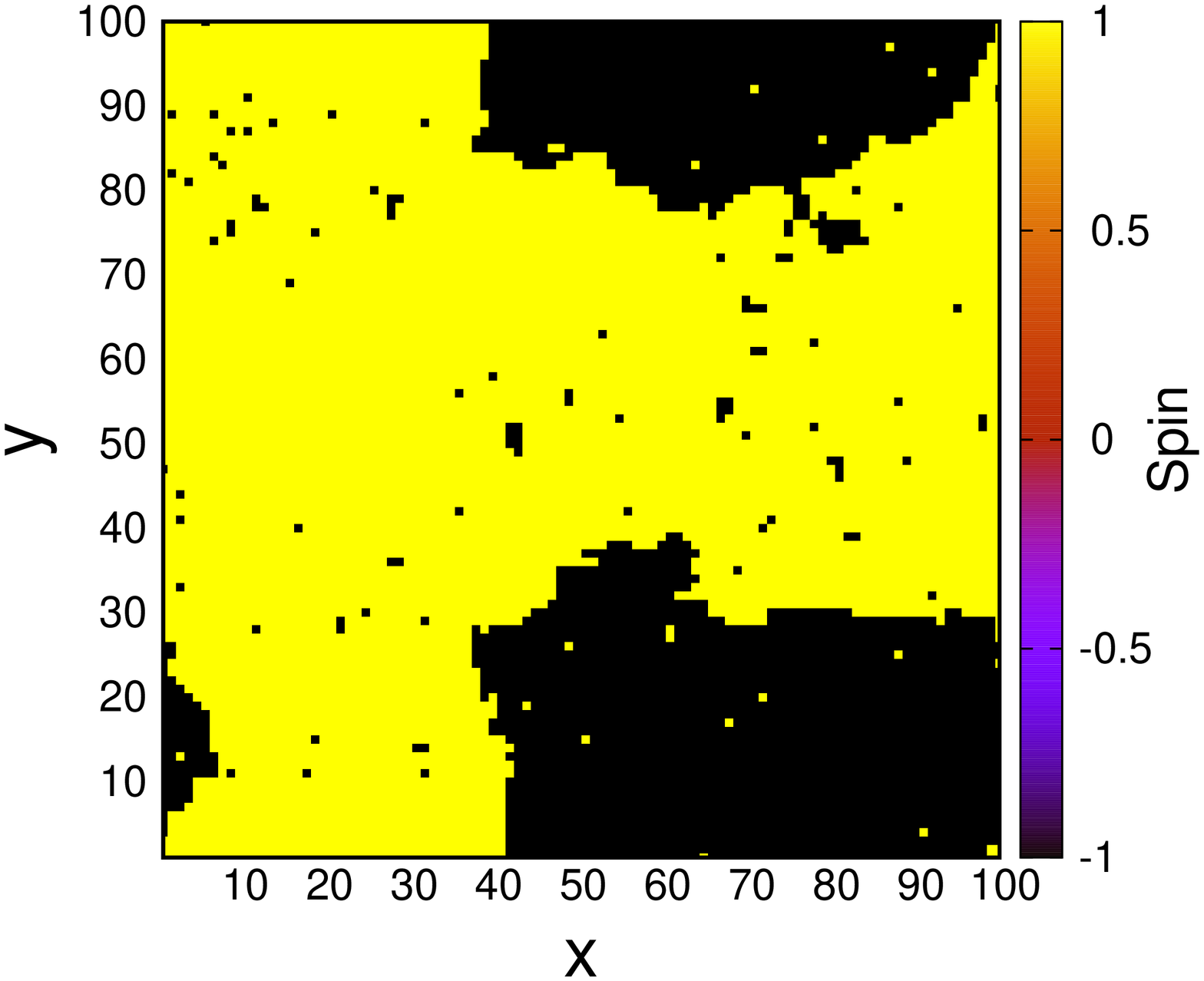}
    \subcaption{}
  \end{subfigure}
  \begin{subfigure}[b]{0.5\textwidth}
    \includegraphics[width=0.7\textwidth,angle=-90]{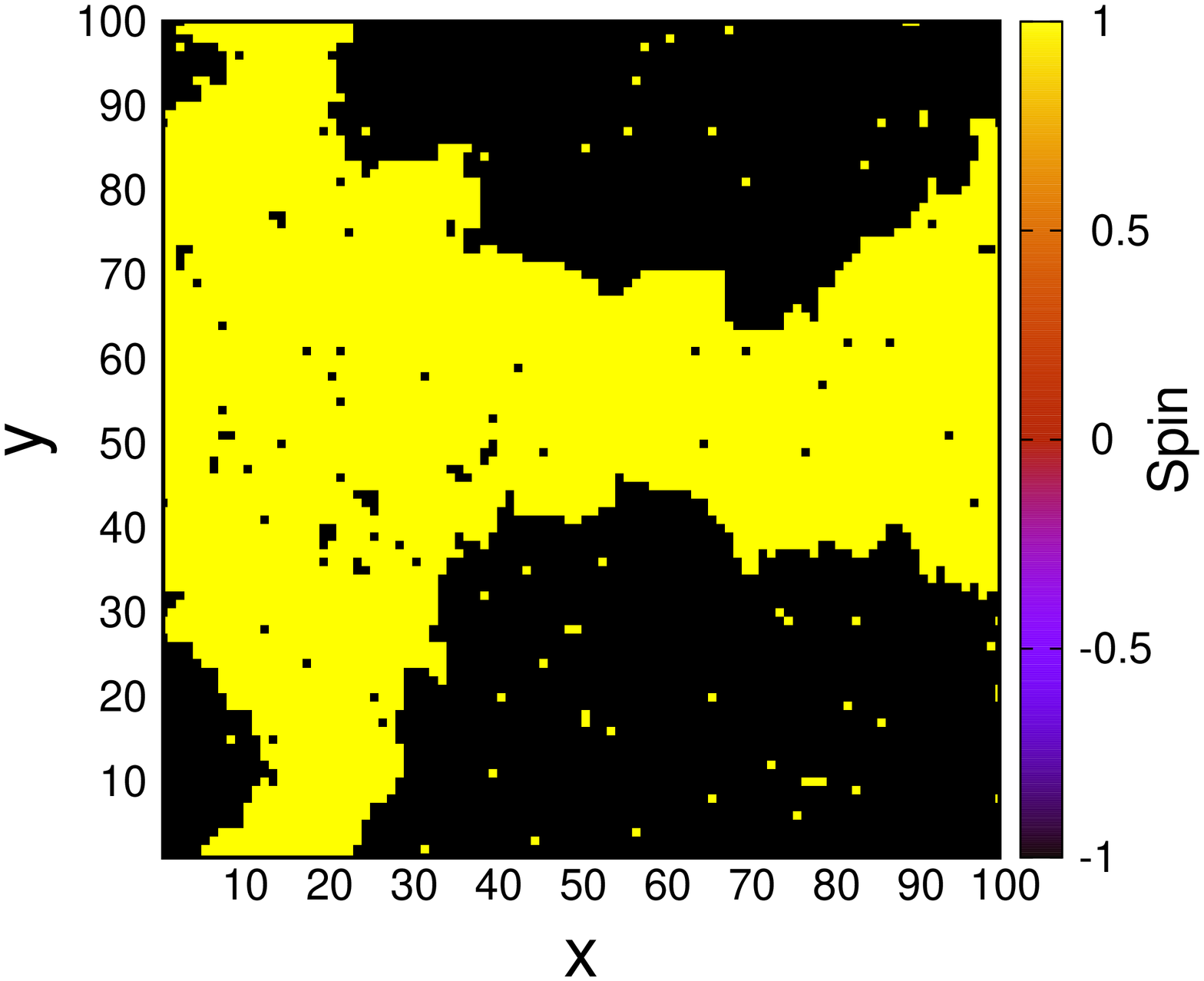}
    \subcaption{}
  \end{subfigure}
  \begin{subfigure}[b]{0.5\textwidth}
    \includegraphics[width=0.7\textwidth,angle=-90]{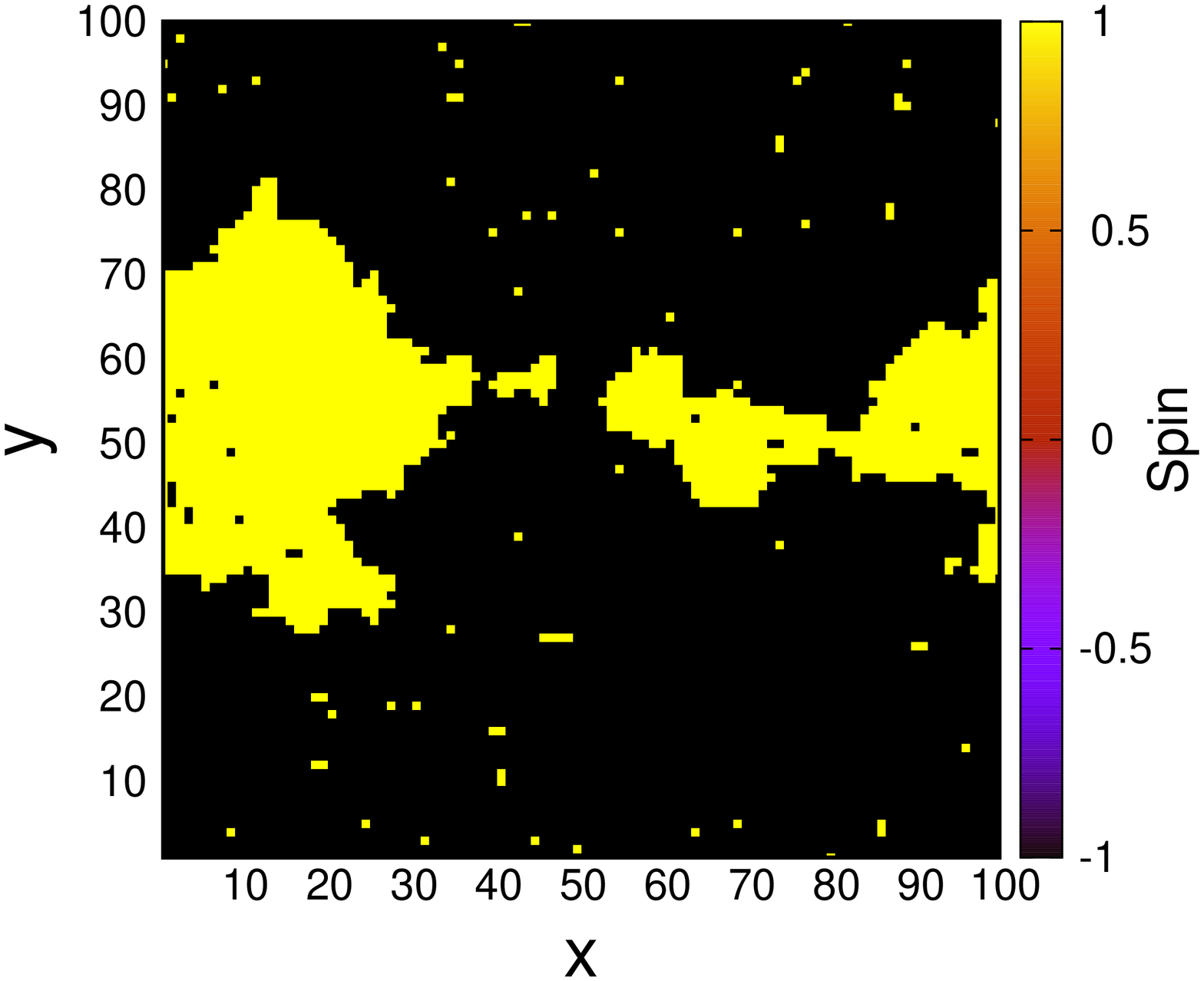}
    \subcaption{}
  \end{subfigure}
\caption{Image plots of the values of the spins at four different Monte Carlo single steps (a)$t=2300$ 
	(before reversal) (b)$t=2400$ (before reversal) (c)$t=2546$ (at reversal) (d)$t=2700$ MCSS 
	(after reversal) in the presence of $h_i$ (bimodal distribution) of width $w=0.3$
	and $h_0=-0.125$.}
\label{snap_bi0.3} 
\end{figure}

\begin{figure}[h!]
  \begin{subfigure}[b]{0.5\textwidth}
    \includegraphics[width=0.7\textwidth,angle=-90]{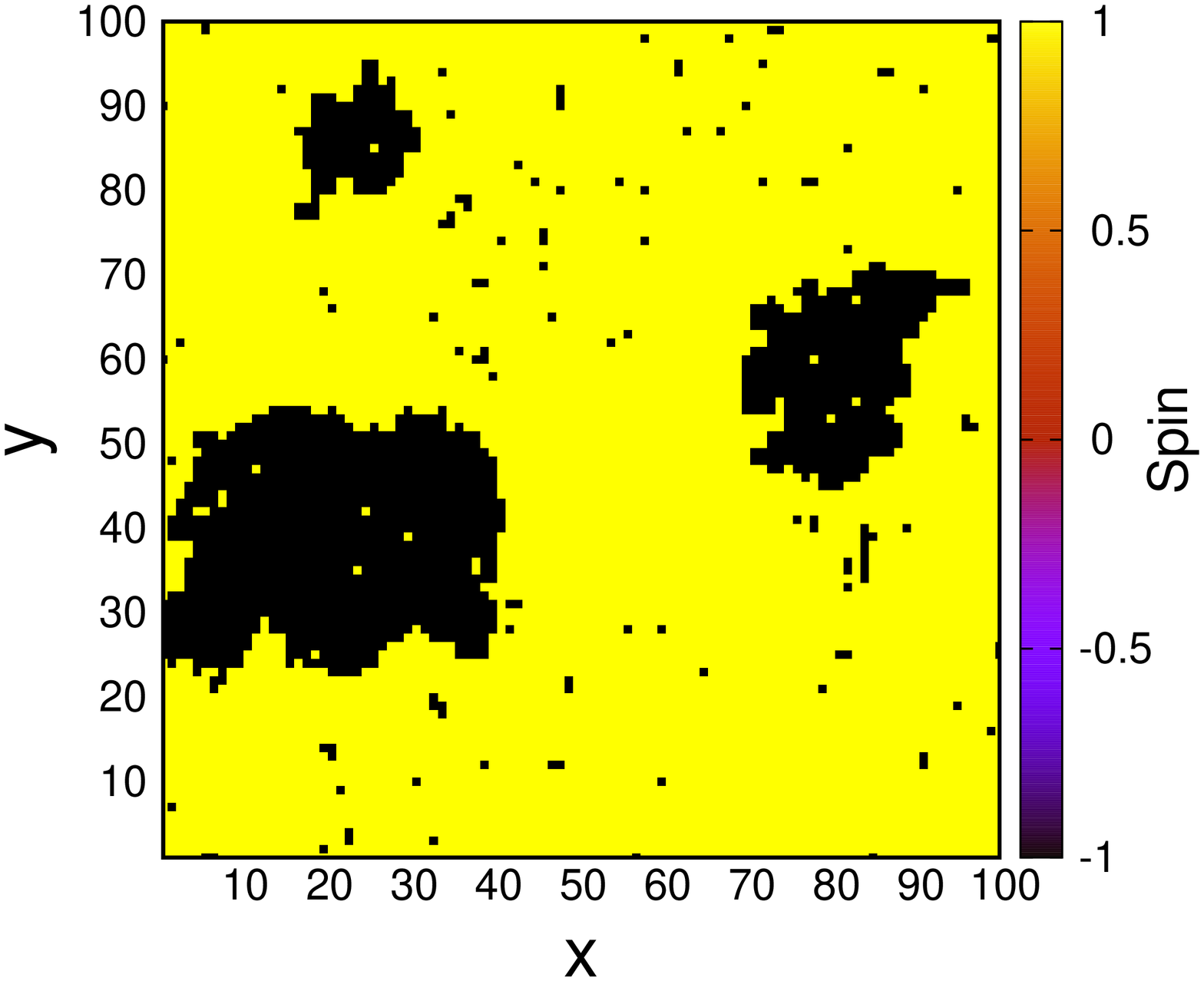}
    \subcaption{}
  \end{subfigure}
  \begin{subfigure}[b]{0.5\textwidth}
    \includegraphics[width=0.7\textwidth,angle=-90]{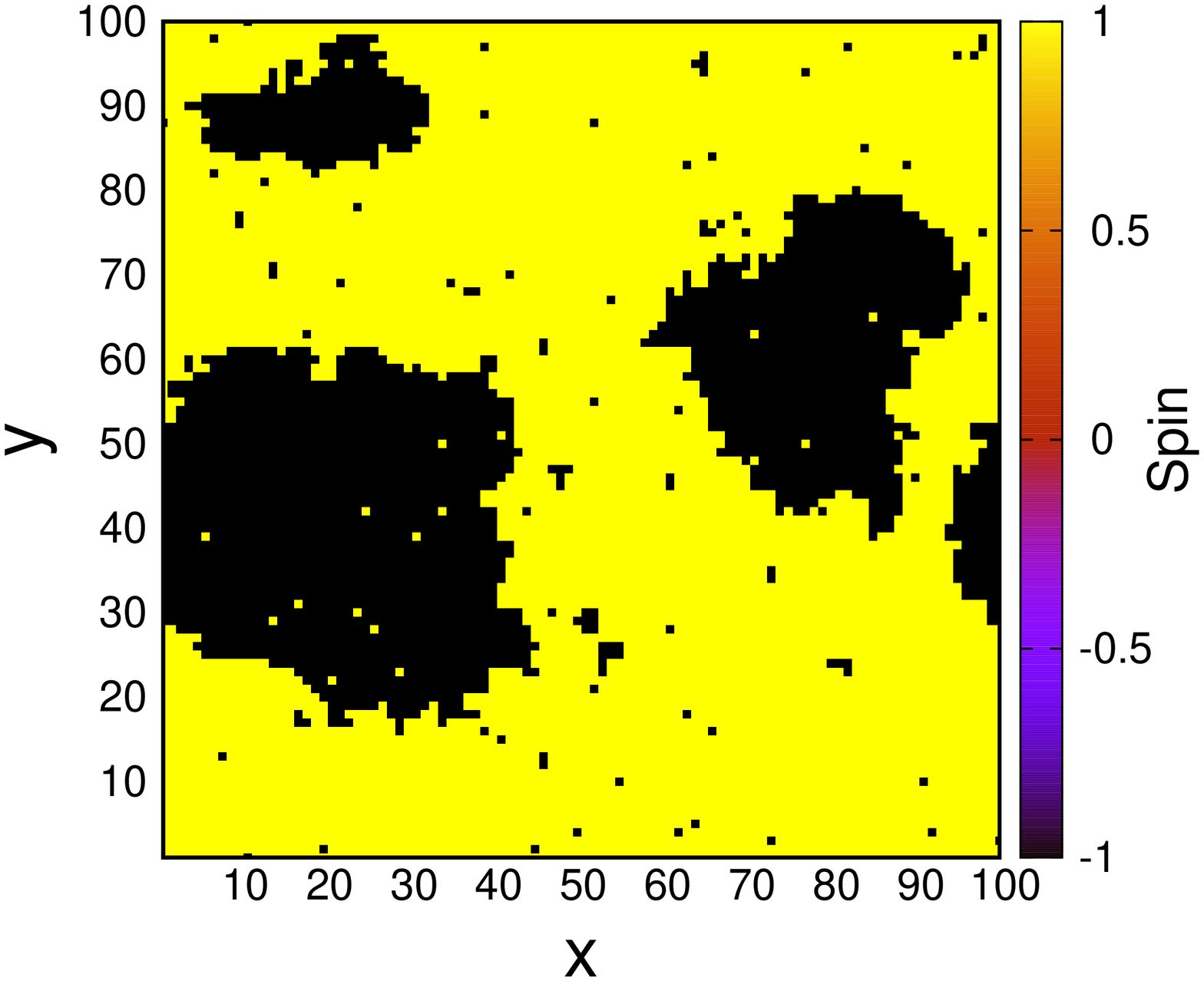}
    \subcaption{}
  \end{subfigure}
  \begin{subfigure}[b]{0.5\textwidth}
    \includegraphics[width=0.7\textwidth,angle=-90]{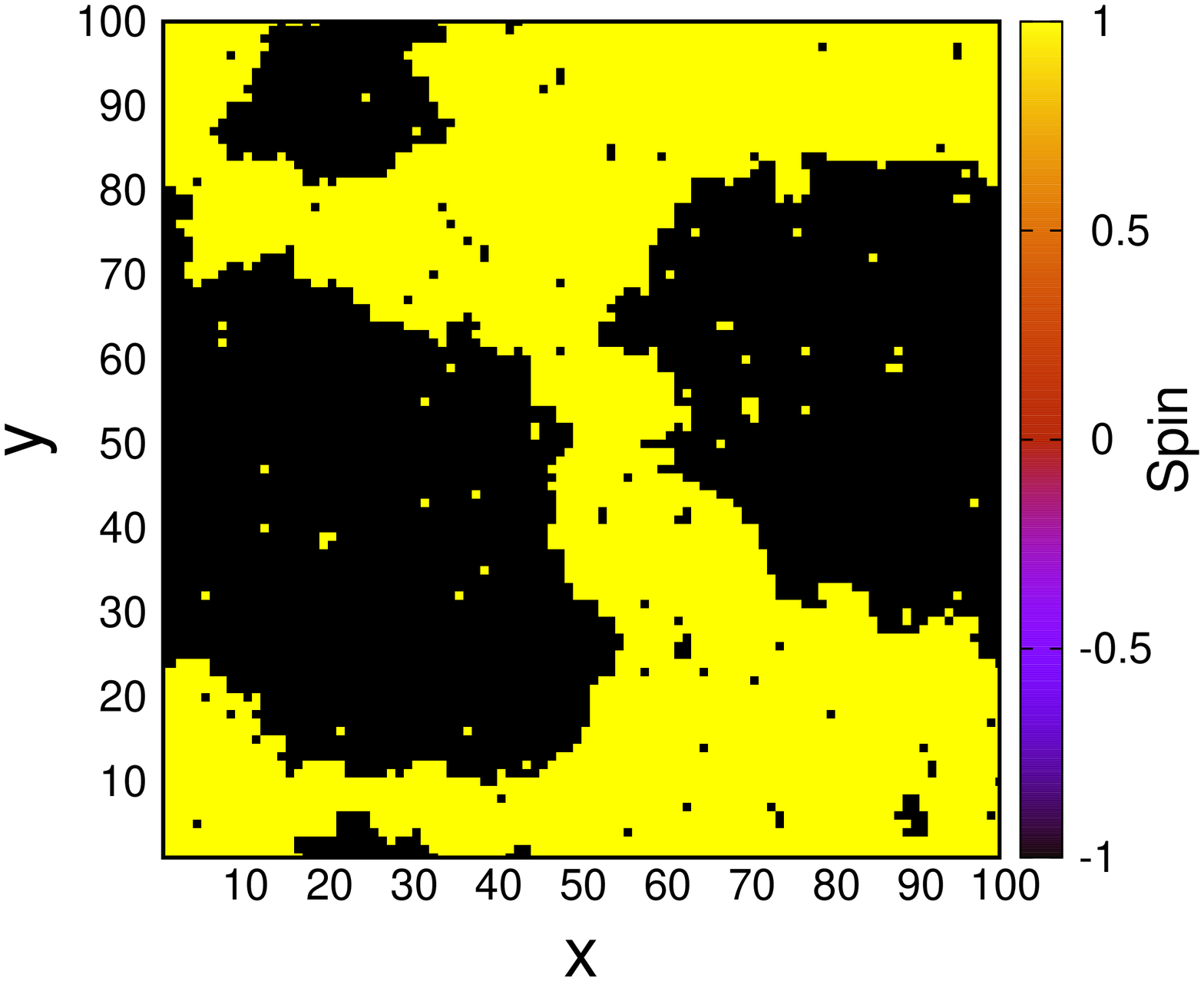}
    \subcaption{}
  \end{subfigure}
  \begin{subfigure}[b]{0.5\textwidth}
    \includegraphics[width=0.7\textwidth,angle=-90]{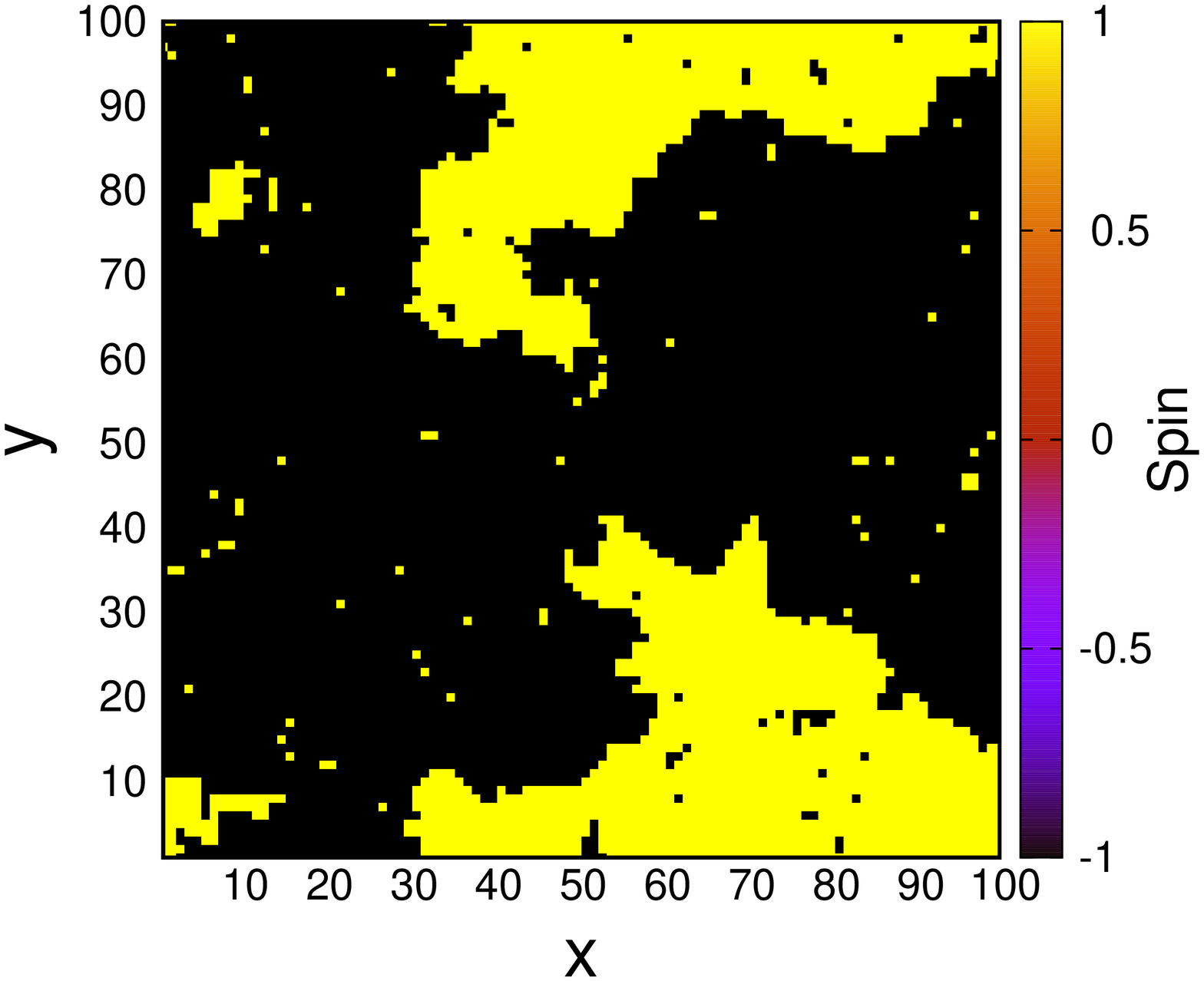}
    \subcaption{}
  \end{subfigure}
\caption{Image plots of the values of the spins at four different Monte Carlo single steps (a)$t=500$ 
	(before reversal) (b)$t=600$ (before reversal) (c)$t=711$ (at reversal) (d)$t=800$ 
	MCSS (after reversal) in the presence of $h_i$ (bimodal distribution) of width $w=0.35$ and $h_0=-0.125$.}
\label{snap_bi0.35}
\end{figure}

In Fig-\ref{snap_bi0.3}, we have taken some snapshots of spin configurations at four different times in the nucleation 
regime i.e. at $|h_0|=0.125$ or $1/|h_0|=8$ in the presence of $h_i$ (bimodal distribution of width $w=0.3$). Here we can see that reversal occurs 
through the formation of a single supercritical droplet growing  with time. Similarly, Fig-\ref{snap_bi0.35} show the snapshots in the presence of bimodal random field having width $w=0.35$. But from 
Fig-\ref{snap_bi0.35} it is clear that in the case of randomly distributed field, the reversal occurs through the growth of many super critical droplets. Nucleation regime exists no more (for $L=100$) in the presence of the bimodal random field distribution of width $w=0.35$. So we can say 
that there must be some limiting value of width of the random field between $w=0.3$ and $w=0.35$ beyond which the nucleation regime no more exists. 

\begin{figure}[h]
\begin{center}
	\resizebox{12cm}{!}{\includegraphics[angle=-90]{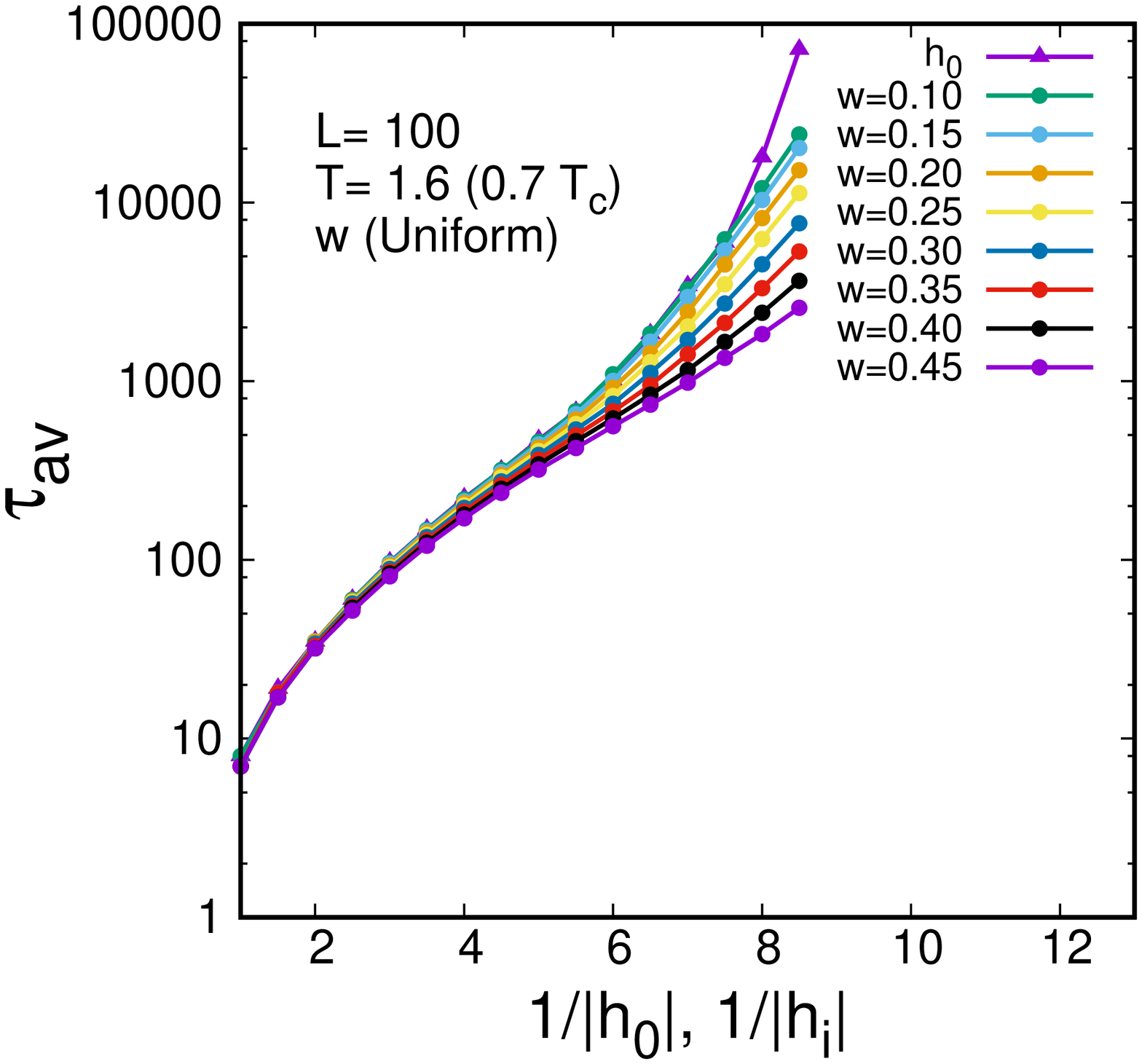}}
	\caption{Variation of mean reversal time (in logarithmic scale) with the inverse of field for lattice size $L=100$ 
		at temperature $T=1.6 (0.7 T_c)$ in the presence of $h_0$ (purple curve 
		with triangular points) and $h_i$ with uniform distribution of different widths $w$ (all other 
		curves with circular points).}
\label{becker_uniran}
\end{center}
\end{figure}

\begin{figure}[h!]
  \begin{subfigure}[b]{0.5\textwidth}
    \includegraphics[width=0.7\textwidth,angle=-90]{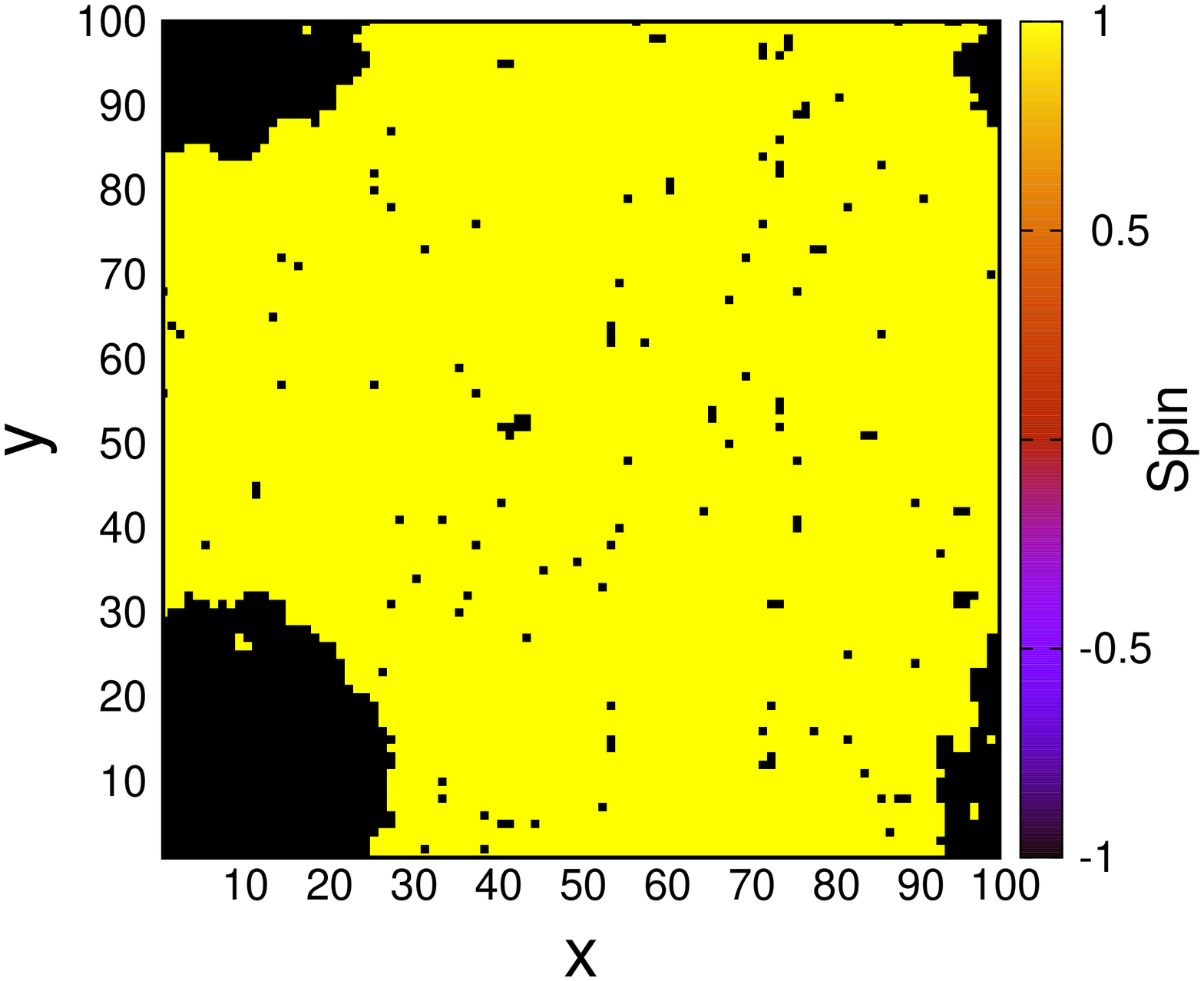}
    \subcaption{}
  \end{subfigure}
  \begin{subfigure}[b]{0.5\textwidth}
    \includegraphics[width=0.7\textwidth,angle=-90]{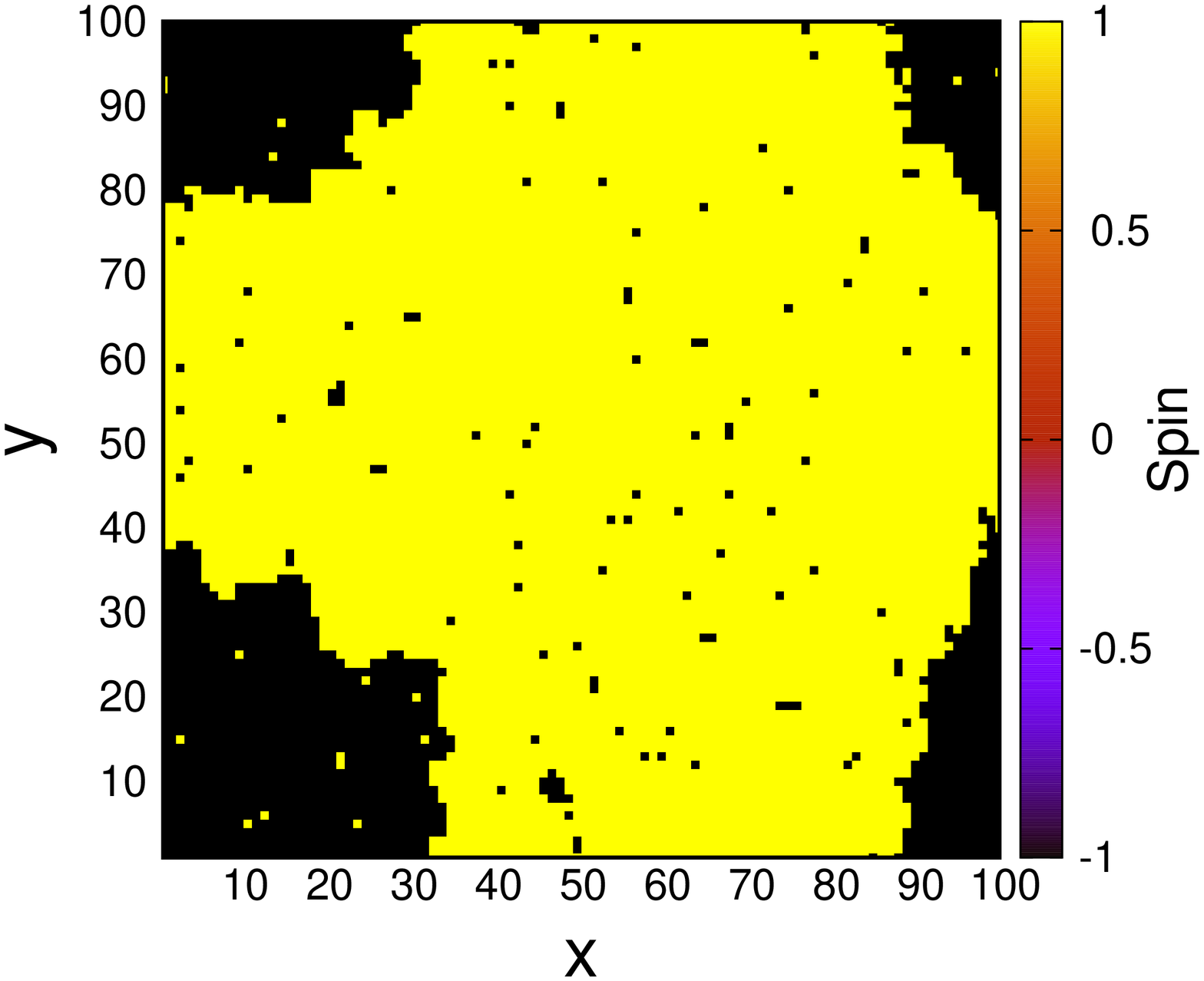}
    \subcaption{}
  \end{subfigure}
  \begin{subfigure}[b]{0.5\textwidth}
    \includegraphics[width=0.7\textwidth,angle=-90]{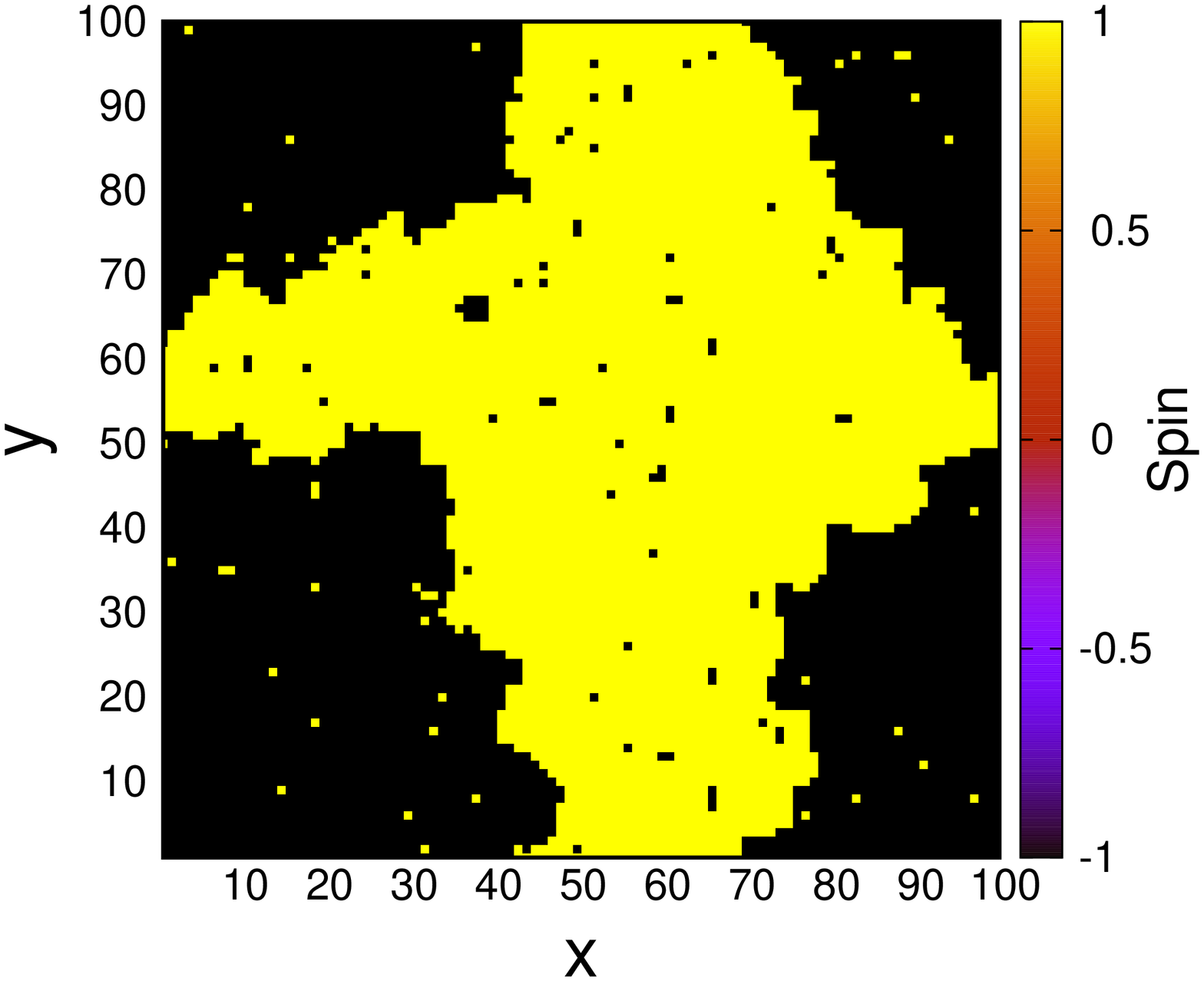}
    \subcaption{}
  \end{subfigure}
  \begin{subfigure}[b]{0.5\textwidth}
    \includegraphics[width=0.7\textwidth,angle=-90]{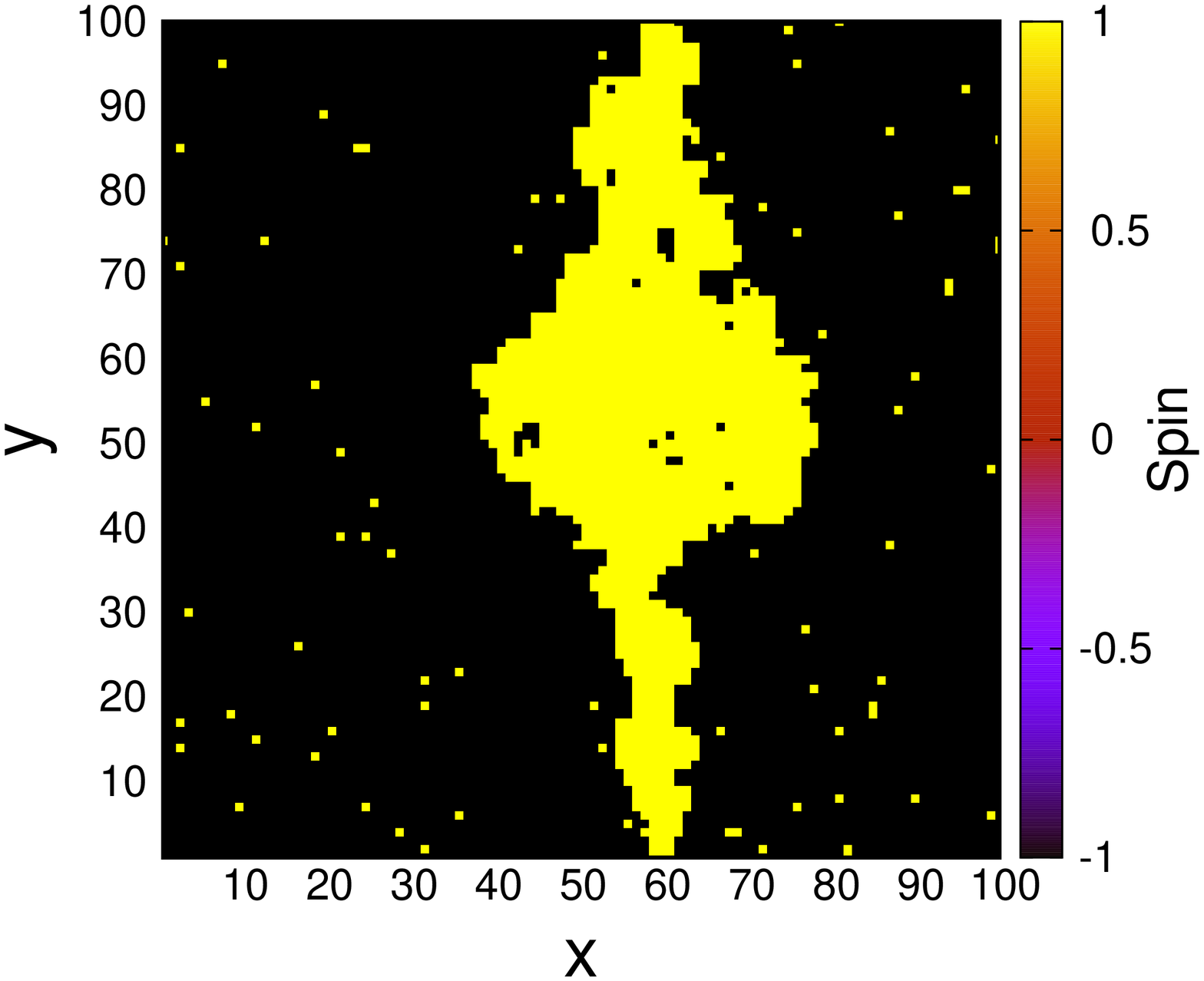}
    \subcaption{}
  \end{subfigure}
\caption{Image plots of the values of the spins at four different Monte Carlo single steps (a)$t=1500$ 
	(before reversal) (b)$t=1600$ (before reversal) (c)$t=1827$ (at reversal) (d)$t=2000$ 
	MCSS (after reversal) in the presence of $h_i$ (uniform distribution) of width $w=0.45$ and $h_0=-0.125$.}
\label{snap_uniran}
\end{figure}

\begin{figure}[h]
\begin{center}
	\resizebox{12cm}{!}{\includegraphics[angle=-90]{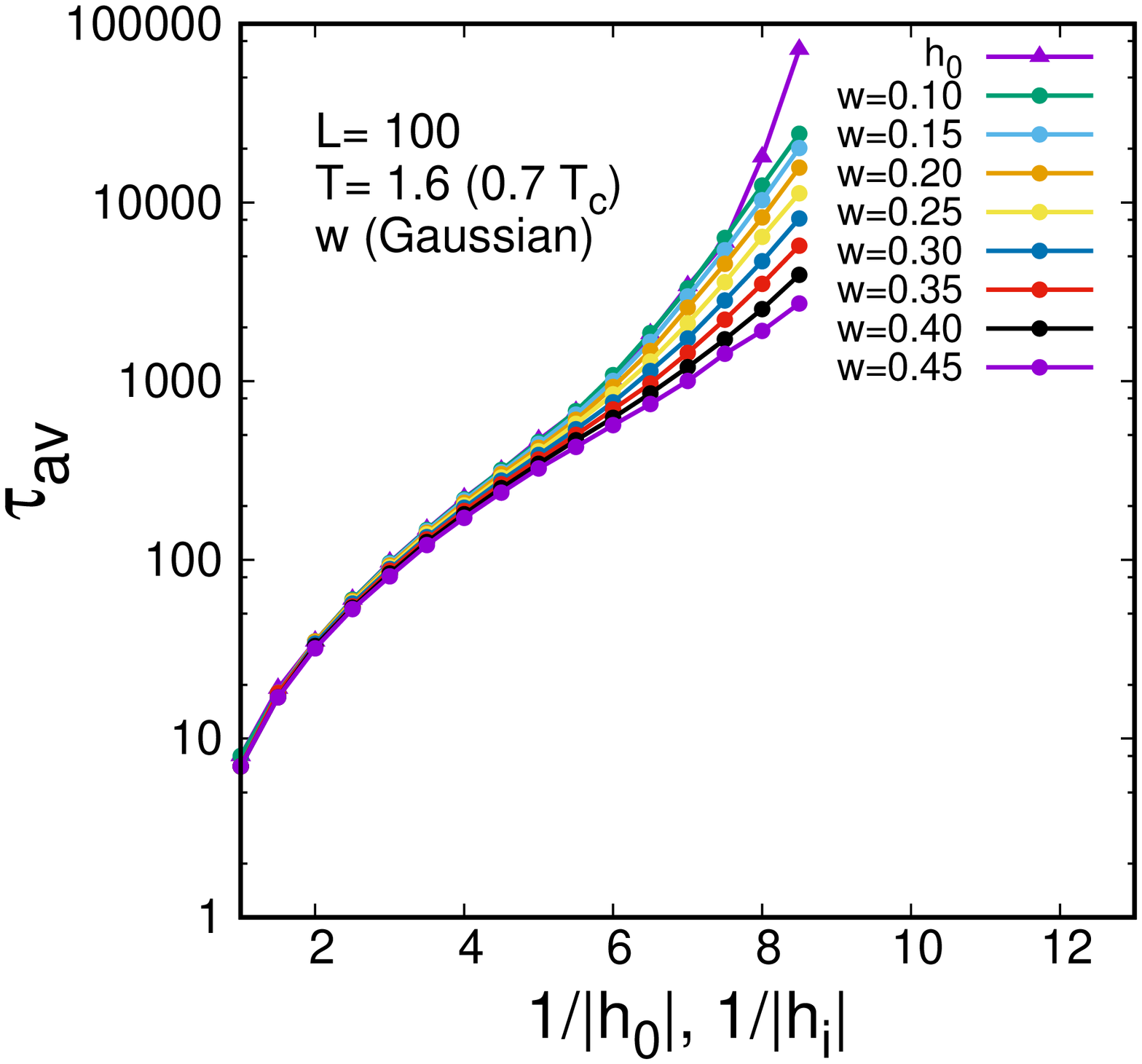}}
	\caption{Variation of mean reversal time (in logarithmic scale) with the inverse of field for lattice size $L=100$ 
		at temperature $T=1.6 (0.7 T_c)$ in the presence of $h_0$ (purple curve 
		with triangular points) and $h_i$ with Gaussian distribution of different widths $w$ (all other 
		curves with circular points).}
\label{becker_gauss}
\end{center}
\end{figure}

\begin{figure}[h!]
  \begin{subfigure}[b]{0.5\textwidth}
    \includegraphics[width=0.7\textwidth,angle=-90]{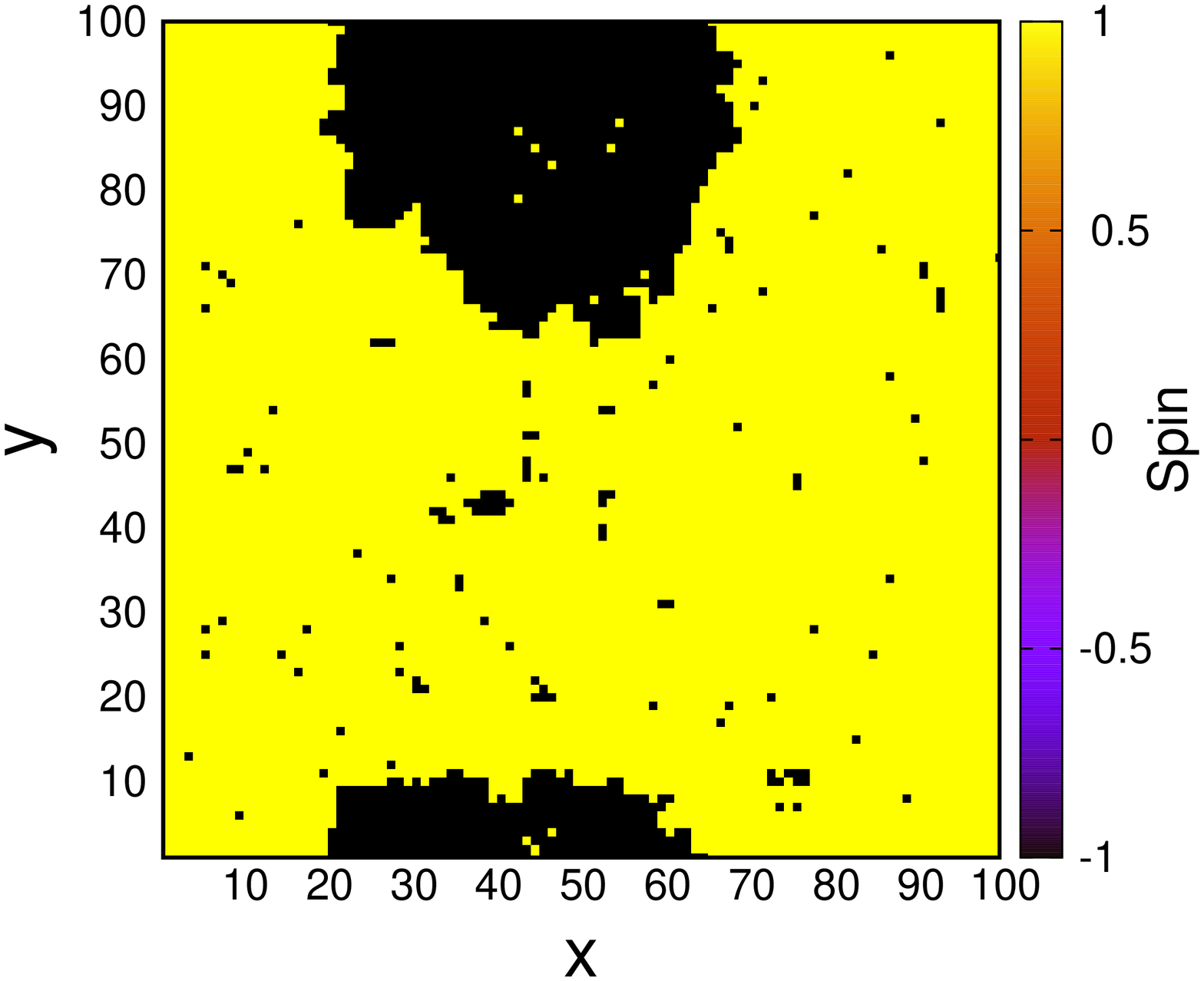}
    \subcaption{}
  \end{subfigure}
  \begin{subfigure}[b]{0.5\textwidth}
    \includegraphics[width=0.7\textwidth,angle=-90]{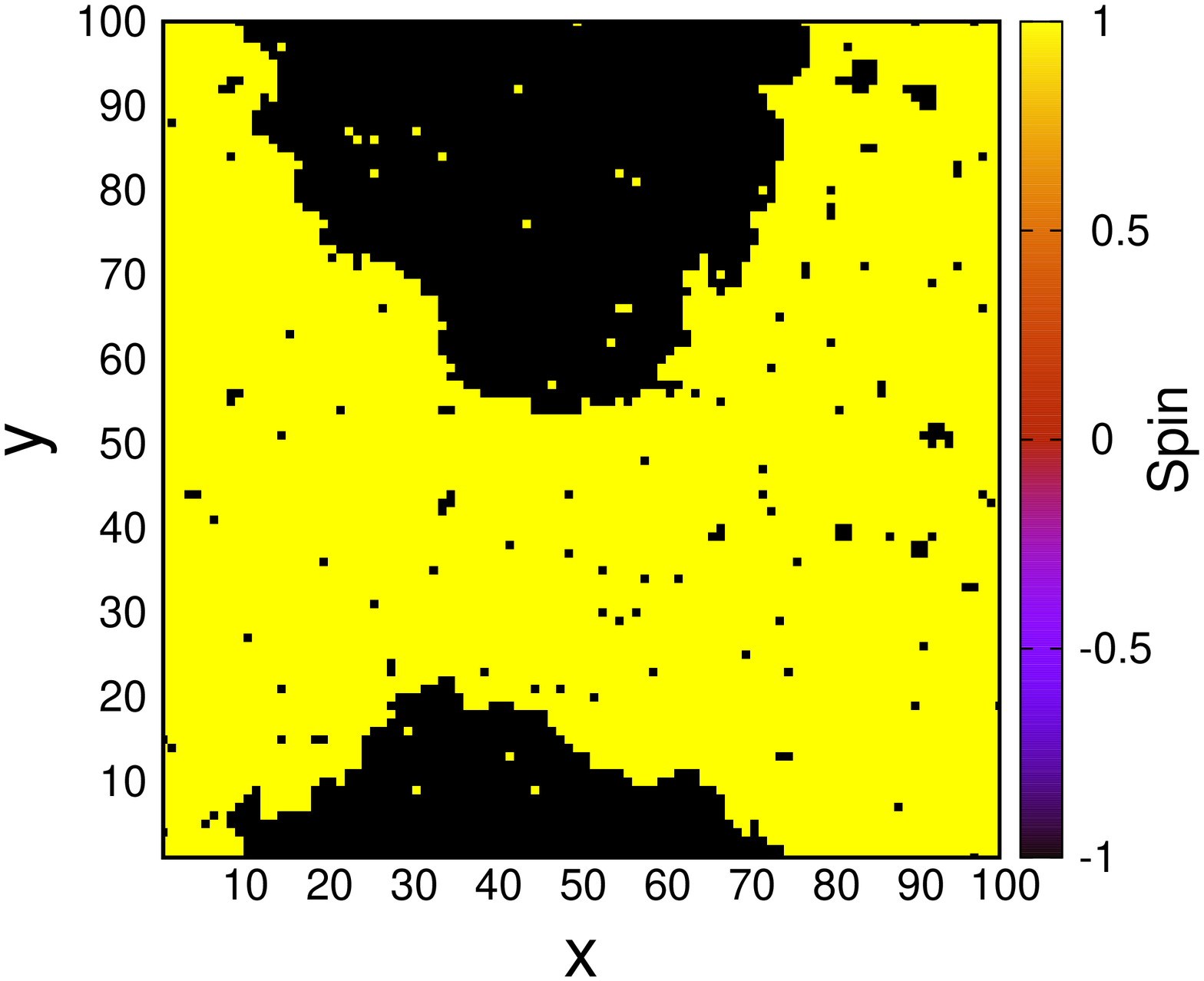}
    \subcaption{}
  \end{subfigure}
  \begin{subfigure}[b]{0.5\textwidth}
    \includegraphics[width=0.7\textwidth,angle=-90]{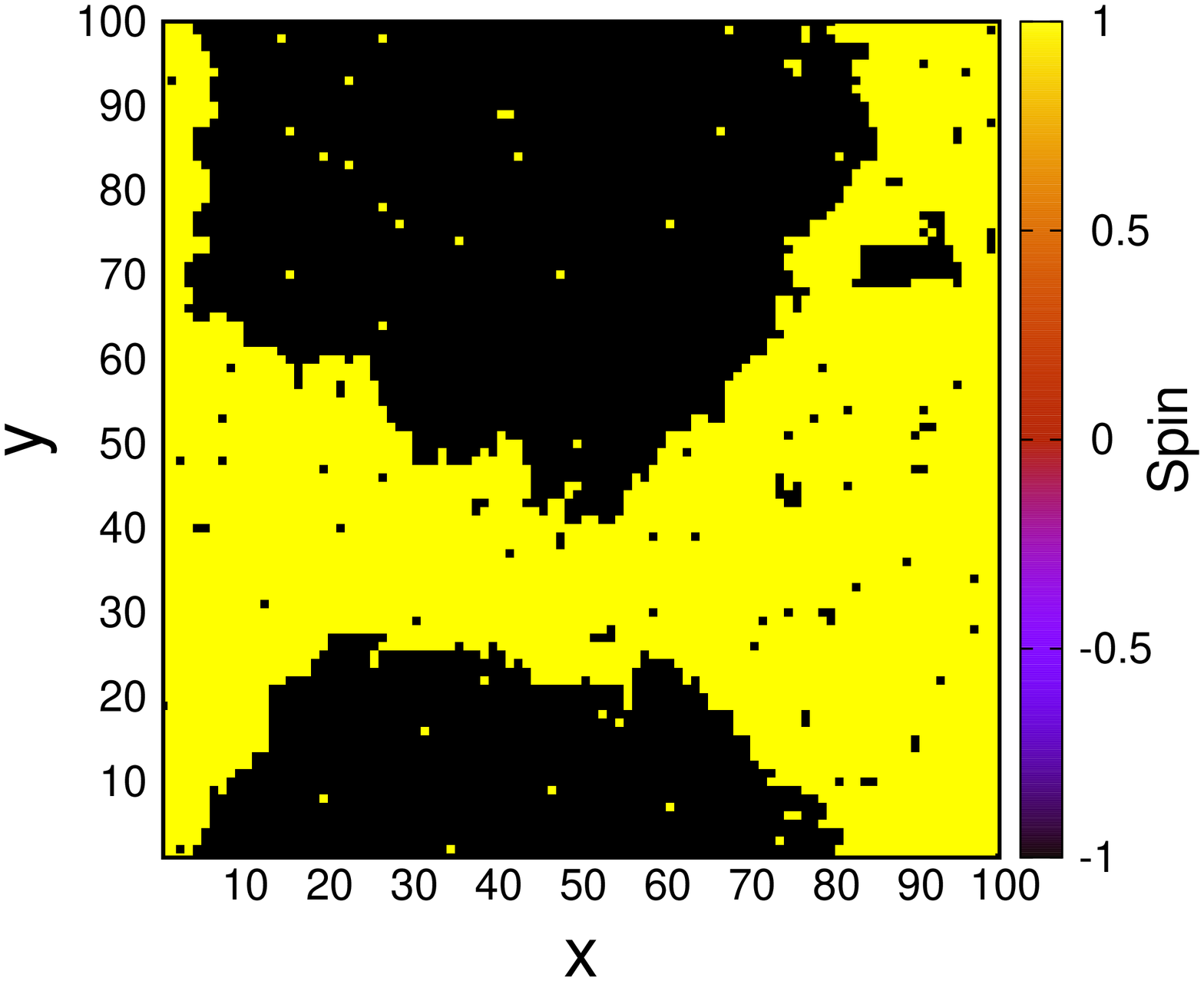}
    \subcaption{}
  \end{subfigure}
  \begin{subfigure}[b]{0.5\textwidth}
    \includegraphics[width=0.7\textwidth,angle=-90]{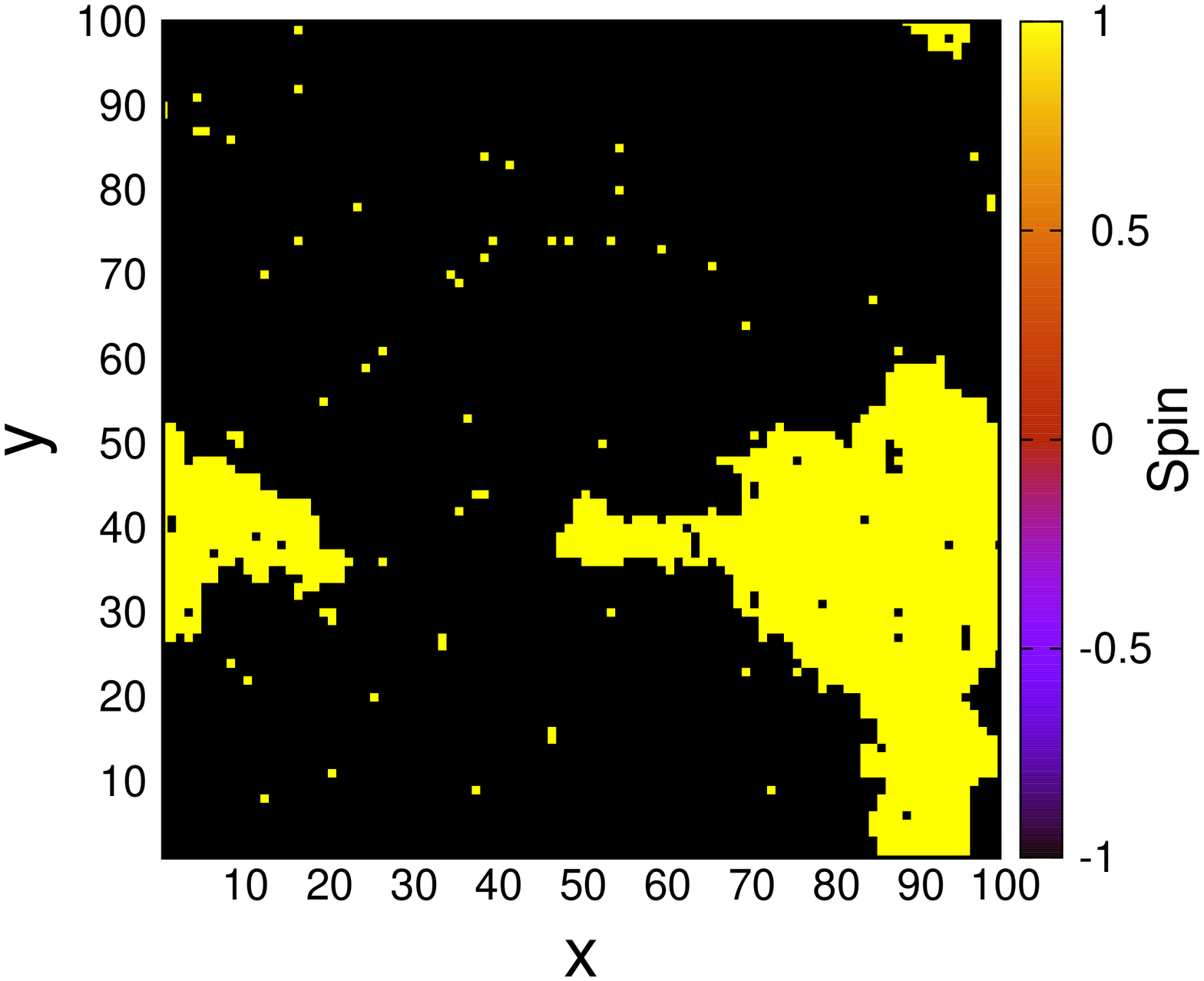}
    \subcaption{}
  \end{subfigure}
\caption{Image plots of the values of the spins at four different Monte Carlo single steps (a)$t=2000$ 
	(before reversal) (b)$t=2100$ (before reversal) (c)$t=2227$ (at reversal) (d)$t=2400$ 
	MCSS (after reversal) in the presence of $h_i$ (Gaussian distribution) of width $w=0.45$ and $h_0=-0.125$.}
\label{snap_gauss}
\end{figure}

Similar studies as above have been carried out in presence of uniform (Fig-\ref{becker_uniran} and Fig-\ref{snap_uniran}) and Gaussian random fields(Fig-\ref{becker_gauss} and Fig-\ref{snap_gauss}). In both case we have observed that the system still remains in nucleation regime in presence of random field of width $w=0.45$. If we take same example as bimodal distribution that, Say the system is in nucleation regime with $h_0= -0.125$ or $\frac{1}{|h_0|} = 8$. Now if we apply a uniform random field ($h_r$) having width $w= 0.3$ i.e. all the values of field between $h_i= -0.125-0.3 = -0.425$ to $h_i= -0.125+0.3 = 0.175$ are present in equal proportion. But among them the lattice sites having the net field values from $h_i= 0$ to $h_i=0.175$ will not participate in the microscopic reversal mechanism. Rest of the sites experiencing the negative values of the fields will take active role in the
microscopic reversal phenomena through the spin flip. Similar fact happens for Gaussian distribution also. So clearly the bimodal distribution is more effective than the other two kinds of distributions of random fields. So there will be some critical width of random field $w>0.45$ beyond which the nucleation regime will disappear in presence of uniform and Gaussian distribution of random field.


\vskip 2 cm
\noindent {\large\bf V. Summary} 
\vskip 0.2 cm

In this paper the behaviour of the metastable state and its persistence has been studied in presence 
of quenched random field by Monte Carlo simulation of spin dynamics on a ferromagnetic Ising square 
lattice. We have studied the change in reversal time of magnetisation by applying three different 
types of the statistical distributions of the random fields. We observed that the reversal time decreases 
in presence of any kind of random fields. Also the reversal time as well as the spin
configurations 
vary with the nature of the distribution of random fields though the mean of each distribution is 
zero. The variation of reversal times with temperature and the width of random fields has been 
studied. The temperature dependence of the reversal time is like $\tau \sim {\rm exp}({A/T})$
(Fig-\ref{temp_var}) and 
the reversal time varies with the width of the distribution of random field is like      $\tau \sim {\rm exp}({-Bw})$
(Fig-\ref{width_var}).
Where $T$ is the temperature of the system, $w$ is the width of the distribution of
the random field and $A$, $B$ are constants. The random field disorder plays a role
similar to that of the temperature.

In the presence of randomly quenched disorder, we have found that the three different regions (namely strong field regime, coalescence regime, and nucleation regime) are
also present as predicted by Becker-D\"oring theory (for uniform field). However, for
stronger (characterised by the width of the distribution) disorder the nucleation
regime (growth of signle droplets) was not observed. Stronger disorder may produce
the net magnetic field positive at some lattice sites which would not help to flip
the spin. So, due to the presence of net positive fields distributed over the entire
lattice would prevent to grow a single droplet. For example, for bimodal distribution
of random field the fifty percent of the lattice sites will produce positive field 
and hence resist to grow a signle droplet to engulf the whole system eventually.


\vskip 2 cm
\noindent {\large\bf V. Acknowledgements}
\vskip 0.3 cm
We would like to thank Soham Chandra and Sk Sajid for useful discussions. MA acknowledges FRPDF research grant provided by the Presidency University. Swami
Vivekanda Scholarship is gratefully acknowledged by MN.
\vskip 0.2 cm


\newpage

\newpage


\newpage

\newpage

\newpage

\newpage

\newpage

\newpage

\newpage

\newpage

\newpage

\newpage


\end{document}